\documentstyle[psfig,twocolumn,floats,eqsecnum,prb,aps]{revtex}
\begin{document}
\draft

\twocolumn[\hsize\textwidth\columnwidth\hsize\csname
@twocolumnfalse\endcsname

\title{Impurity-Induced Quasiparticle Transport and Universal Limit
Wiedemann-Franz Violation in d-Wave Superconductors}
\author{Adam C. Durst and Patrick A. Lee}
\address{Department of Physics, Massachusetts Institute of Technology,
Cambridge, Massachusetts 02139}
\date{\today}

\maketitle

\begin{abstract}
Due to the node structure of the gap in a d-wave superconductor, the presence
of impurities generates a finite density of quasiparticle excitations at
zero temperature.  Since these impurity-induced quasiparticles are both
generated and scattered by impurities, prior calculations indicate
a universal limit ($\Omega\rightarrow0$,$T\rightarrow0$) where the transport
coefficients obtain scattering-independent values, depending only on the
velocity anisotropy $v_{f}/v_{2}$.  We improve upon prior results, including
the contributions of vertex corrections and Fermi liquid corrections in our
calculations of universal limit electrical, thermal, and spin conductivity.
We find that while vertex corrections modify electrical conductivity and
Fermi liquid corrections renormalize both electrical and spin conductivity,
only thermal conductivity maintains its universal value, independent of
impurity scattering or Fermi liquid interactions.  Hence, low temperature
thermal conductivity measurements provide the most direct means of obtaining
the velocity anisotropy for high $T_{c}$ cuprate superconductors.
\end{abstract}

\pacs{PACS numbers: 74.25.Fy, 74.72.-h, 71.55.-i}

]

\section{Introduction}
\label{sec:intro}
The characteristic feature of a d-wave superconductor is the existence of
four nodal points where the order parameter vanishes.  Since low energy
excitations are concentrated about these nodes, low temperature behavior is
dominated by the details of the node structure and, in particular, the ratio
of the Fermi velocity to the gap velocity (slope), $v_{f}/v_{2}$.  Prior
theoretical work has shown that this velocity ratio is prominent in
expressions for low temperature transport coefficients 
\cite{lee93,hir93,hir94,hir96,gra96,sen98,bal94} as well as the
temperature dependence of the superfluid density \cite{lee97,mil98,xu95}.
However, discrepancies between values of $v_{f}/v_{2}$ obtained from
measurements of microwave electrical conductivity \cite{hos98},
thermal conductivity \cite{tai97,chi99}, and superfluid density
\cite{bon96}, as well as direct measurements of gap structure via
ARPES \cite{mes99} indicate that the existing theoretical predictions
must be corrected through a more detailed analysis.  To this end, we
calculate herein electrical, thermal, and spin conductivity including the
contributions of vertex corrections and Fermi liquid corrections.
Associated calculations of the superfluid density will be pursued in a
future investigation.

It has been shown \cite{gor85} that for a superconductor with
$d_{x^{2}-y^{2}}$ pairing symmetry, the presence of impurities generates
a finite density of quasiparticle states down to zero energy
(although the ultra-low energy regime remains the subject of some
debate \cite{ner95,sen98A}).  This
results in a unique situation where an increase in impurity density
increases the density of quasiparticles while reducing the quasiparticle
lifetime.  As a result of the cancellation of these opposing effects,
``bare bubble'' conductivity calculations (neglecting the corrections
we shall consider) indicate a universal limit
($\Omega\rightarrow0$,$T\rightarrow0$) where the transport coefficients
attain constant values, independent of scattering. \cite{lee93}
However, we shall see that these results are modified by two types of
corrections: vertex corrections and Fermi liquid corrections.  Vertex
corrections account for the fact that forward scattering does not
interfere with the progress of a carrier to the same extent as back
scattering.  Hence, if the scattering potential varies in k-space such
that the potential for forward scattering differs from that for back
scattering, the bare bubble transport coefficients may be modified.
Fermi liquid corrections account for the underlying Fermi liquid
interactions between electrons in the superconductor.  Due to such
interactions, the presence of a quasiparticle current induces an
additional drag current which may renormalize the transport
coefficients.  The purpose of what follows is to improve upon the
bare bubble results by including the effects of both types of corrections.

In Sec.\ \ref{sec:dos}, we define the parameters of our phenomenological
d-wave model, introduce the Green's function, and calculate the density
of states.  In Appendix \ref{app:barebubble}, neglecting all corrections,
we derive a generalized bare bubble polarization function which can be
applied to the calculation of either electrical, thermal, or spin
conductivity.  By treating the general case, we avoid repeating the same
basic calculation three times.  In Appendix \ref{app:vertexcorrections},
we calculate another generalized polarization function, now including the
contributions of vertex corrections.  The significance of vertex corrections
in the universal limit is determined via a numerical calculation presented
in Appendix \ref{app:numerical}.  In Appendix \ref{app:fermiliquid},
we derive the renormalization of a generalized current due to the
effects of underlying Fermi liquid interactions.  In Secs.\
\ref{sec:elec}, \ref{sec:therm}, and \ref{sec:spin}, we make use of the
results in the appendices to calculate electrical, thermal, and spin
conductivity in the universal limit ($\Omega\rightarrow0$,$T\rightarrow0$).
Each of these sections begins with a derivation of the appropriate 
current density operator.  These calculations reveal an extra gap velocity
term in the thermal and spin currents due to the momentum dependence of
the d-wave gap and therefore indicate a correction to the standard
thermal conductivity formula \cite{amb65,amb64} derived assuming
an s-wave gap.  Given each current operator, 
we present the bare bubble result and
then note modifications due to vertex corrections and Fermi liquid
corrections.  We find that contrary to the scattering-independent result
obtained from the bare bubble calculation, the universal limit electrical
conductivity attains a vertex correction, which depends explicitly on the
nature of the impurity scattering, as well as a Fermi liquid renormalization,
which depends on the strength of Fermi liquid interactions.  In addition, while
the spin conductivity is unaffected by vertex corrections (for small impurity
density), it is renormalized due to Fermi liquid interactions.  Only the 
thermal conductivity has neither a vertex correction nor a Fermi liquid
correction.  It therefore retains its simple, universal value.  Conclusions are
discussed in Sec.\ \ref{sec:conclusions} where we provide physical descriptions
of the mathematical corrections calculated herein.

\section{D-Wave Model, Green's Function, \protect\\ and Density of States}
\label{sec:dos}
To study the low temperature transport properties of a d-wave superconductor,
we employ a phenomenological model \cite{lee93,lee97} with the
Brillouin zone of a two-dimensional square lattice (of lattice constant $a$),
an electron dispersion (via tight binding parametrization)
\begin{equation}
\epsilon_{k} = -2 t_{f} (\cos k_{x}a + \cos k_{y}a) - \mu ,
\label{eq:dispersiondef}
\end{equation}
and an order parameter of $d_{x^{2}-y^{2}}$ symmetry
\begin{equation}
\Delta_{k} = \frac{\Delta_{0}}{2} (\cos k_{x}a - \cos k_{y}a)
\label{eq:deltadef}
\end{equation}
which crosses through zero at each of four nodal points on the Fermi surface,
$(k_{x}= \pm k_{y})$.
The key feature of such a model is that in the vicinity of each of the gap
nodes, $\epsilon_{k}$ varies linearly across the Fermi surface
and $\Delta_{k}$ varies linearly along the Fermi surface.  Defining local
momentum variables at each of the nodes with $\hat{\bf k}_{1}$ perpendicular
to the Fermi surface and $\hat{\bf k}_{2}$ parallel to the Fermi surface,
we can designate at each node both a Fermi velocity
\begin{equation}
{\bf v}_{f} \equiv \frac{\partial \epsilon_{k}}{\partial {\bf k}}
= v_{f} \hat{\bf k}_{1} \;\;\;\;\;\;\;\; v_{f} = 2 \sqrt{2}\, t_{f} a
\label{eq:velfdef}
\end{equation}
and a gap velocity
\begin{equation}
{\bf v}_{2} \equiv \frac{\partial \Delta_{k}}{\partial {\bf k}}
= v_{2} \hat{\bf k}_{2} \;\;\;\;\;\;\;\; v_{2} = \frac{1}{\sqrt{2}} \Delta_{0} a .
\label{eq:vel2def}
\end{equation}
(Note that all velocities in our model are taken to be ``renormalized'' velocities
accounting for both band structure and many-body effects
within the context of Fermi liquid theory.)
Utilizing these definitions, it becomes clear that the quasiparticle excitation
spectrum in the vicinity of each of the gap nodes takes the form of an
anisotropic Dirac cone
\begin{equation}
E_{k} = \sqrt{\epsilon_{k}^{2} + \Delta_{k}^{2}}
= \sqrt{v_{f}^{2} k_{1}^{2} + v_{2}^{2} k_{2}^{2}}
\label{eq:diracspectrum}
\end{equation}
where the degree of anisotropy is measured by the ratio of the two velocities.
This ratio $v_{f}/v_{2}$ appears prominently in the low temperature transport
coefficients and is a measurable quantity which provides
a convenient means of comparing theory to experiment.

The low temperature physics of such a model is driven by the
fact that at the four nodes, there is no gap to quasiparticle excitations.  Hence,
quasiparticles are generated only in the vicinity of the gap nodes.  This is very useful
mathematically since it means that a momentum integral over the Brillouin zone
can usually be replaced by a sum over nodes and an integral over the small region
of k-space surrounding each node.  Furthermore, due to the form of the excitation 
spectrum (\ref{eq:diracspectrum}), it is convenient to scale out the anisotropy of
the Dirac cone, and change to polar coordinates in a new scaled momentum
${\bf p}=(p,\theta)$.  Hence we will frequently make the substitution
\begin{equation}
\sum_{k} \rightarrow \sum_{j=1}^{4}\int\frac{dk_{1}dk_{2}}{(2\pi)^{2}}
\rightarrow \sum_{j=1}^{4}\int_{0}^{p_{0}}\frac{p\, dp}{2 \pi v_{f} v_{2}}
\int_{0}^{2\pi}\frac{d\theta}{2\pi}
\label{eq:substitution}
\end{equation}
where $p_{1}=v_{f}k_{1}=p\cos\theta$, $p_{2}=v_{2}k_{2}=p\sin\theta$,
$p=\sqrt{p_{1}^{2}+p_{2}^{2}}=E_{k}$, and
$p_{0}=\sqrt{\pi v_{f}v_{2}}/a \sim {\cal O}(\Delta_{0})$ is a large scaled
momentum cutoff defined such that the area of the new integration region is the
same as that of the original Brillouin zone.  Note that if
quasiparticles are only generated at the nodes and the rest of the Brillouin zone
makes no contribution, then we should be safe in extending this limit to infinity.
However, it is sometimes necessary to retain $p_{0}$ through the intermediate
stages of a calculation (usually as part of a ratio within a logarithm) maintaining
throughout that all other energies are much smaller than this cutoff value.

\begin{figure}[tb]
\centerline{\psfig{file=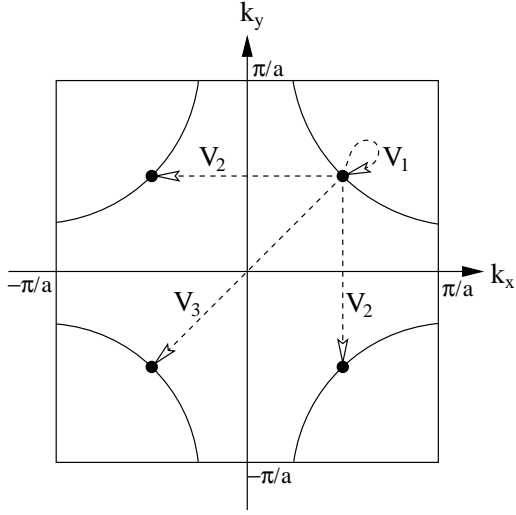}}
\vspace{0.5 cm}
\caption{Impurity scattering within d-wave model.
$V_{1}$, $V_{2}$, and $V_{3}$ are the
potentials for intra-node, adjacent-node, and opposite-node scattering.}
\label{fig:scatteringpotential}
\end{figure}

The fact that quasiparticles are concentrated in the vicinity of the gap nodes
is also very useful when considering impurity scattering.  Since the 
initial and final momenta of a scattering event must always be approximately
equal to the k-space location of one of the four nodes, a general scattering
potential, $V_{kk'}$, need only be evaluated in three possible cases:
intra-node scattering (${\bf k}$ and ${\bf k}'$ at the same node), 
adjacent-node scattering (${\bf k}$ and ${\bf k}'$ at adjacent nodes), and
opposite-node scattering (${\bf k}$ and ${\bf k}'$ at opposite nodes).
These are depicted graphically in Fig.\ \ref{fig:scatteringpotential} and
denoted respectively as $V_{1}$, $V_{2}$, and $V_{3}$.
Hence, an arbitrary potential (varying slowly over the area of a node)
is effectively reduced to a set of three parameters.  This simplification
proves quite helpful.

Since the transport calculations in the sections that follow consist of the
evaluation of Feynman diagrams via field-theoretic techniques, it is important
to establish the types of Green's functions that will be utilized.  For any
superconductor, the existence of a condensate of ground state pairs means
that the annihilation of an electron must be treated on the same footing as
the creation of its mate, an electron with opposite momentum and spin.
Hence we use the Nambu formalism \cite{sch64} in which the field operators are
two-component spinors of the form
\begin{equation}
\Psi_{k} = \left( \begin{array}{c} c_{k\uparrow} \\
c^{\dagger}_{-k\downarrow} \end{array} \right) \;\;\;\;\;\;\;\;
\Psi^{\dagger}_{k} = \left( c^{\dagger}_{k\uparrow} , c_{-k\downarrow} \right)
\label{eq:NambuPsidef}
\end{equation}
and the resulting Green's functions are 2x2 matrices in
{\em Nambu space}.  Since we are concerned with finite temperature
calculations, all diagrams will be evaluated using
the Matsubara finite temperature formalism \cite{mah90}.
Hence, the bare Matsubara Green's function
expressed in Nambu formalism takes the form
\begin{equation}
\tilde{\cal G}_{0}({\bf k},i\omega) = \frac{1}{(i\omega)^{2} - E_{k}^{2}}
\left( \begin{array}{cc} i\omega + \epsilon_{k} & \Delta_{k} \\
\Delta_{k} & i\omega - \epsilon_{k} \end{array} \right)
\label{eq:bareNambuMatGreendef}
\end{equation}
where the tilde denotes a Nambu space matrix and $i\omega=i(2n+1)\pi/\beta$
is a fermionic Matsubara frequency.  In the presence of impurities, the
bare Green's function is dressed via scattering from the impurities and
obtains a Matsubara self-energy, $\tilde{\Sigma}(i\omega)$.  Assuming that
all but the scalar component of the self-energy can be neglected or
absorbed into $\epsilon_{k}$ or $\Delta_{k}$, Dyson's equation yields
that the dressed Matsubara Green's function is given by
\begin{eqnarray}
\tilde{\cal G}({\bf k},i\omega)
&=& \frac{1}{(i\omega - \Sigma(i\omega))^{2} - E_{k}^{2}} \nonumber \\
&\times& \left( \begin{array}{cc} i\omega - \Sigma(i\omega) + \epsilon_{k}
& \Delta_{k} \\ \Delta_{k}
& i\omega - \Sigma(i\omega) - \epsilon_{k} \end{array} \right) .
\label{eq:NambuMatGreendef}
\end{eqnarray}
(Note that while this assumption has been explicitly justified in both the
Born and unitary scattering limits, the omitted self-energy components can
contribute for arbitrary scattering. \cite{mon87,hir88}
For simplicity, we neglect such contributions in this investigation.)
From the Matsubara functions, corresponding retarded functions are obtained
by analytically continuing $i\omega\rightarrow\omega+i\delta$ such that
\begin{equation}
\tilde{G}_{ret}({\bf k},\omega) = \tilde{\cal G}({\bf k},i\omega \rightarrow
\omega+i\delta)
\label{eq:Greenretdef}
\end{equation}
and the impurity scattering rate is defined as
\begin{equation}
\Gamma(\omega) = -\mbox{Im}\, \Sigma_{ret}(\omega)
\label{eq:scatratedef}
\end{equation}
where $\Sigma_{ret}(\omega) = \Sigma(i\omega \rightarrow \omega+i\delta)$.

With the Green's function in hand, it is a simple and illustrative step to
calculate the density of states.  In terms of the retarded Green's function,
the density of states is given by
\begin{equation}
N(\omega) = -\frac{1}{2\pi} \sum_{k} \mbox{Tr} \left[ \mbox{Im}\,
\tilde{G}_{ret}({\bf k},\omega) \right] .
\label{eq:dosdef}
\end{equation}
Plugging in the Green's function (\ref{eq:NambuMatGreendef}),
replacing the sum by a scaled integral about each node via
(\ref{eq:substitution}), neglecting the real part of the self-energy,
and performing the integration we find that
\begin{eqnarray}
N(\omega) &=& \frac{2}{\pi^{2} v_{f} v_{2}} \Gamma(\omega) \left[
\ln\frac{p_{0}}{\Gamma(\omega)} - \ln\sqrt{1+\frac{\omega^{2}}{\Gamma(\omega)^{2}}}
\right] \nonumber \\
&+& \frac{|\omega|}{\pi v_{f} v_{2}} \left[ \frac{1}{2} - \frac{1}{\pi}
\arctan \left( \frac{\Gamma(\omega)^{2} - \omega^{2}}{2 |\omega| \Gamma(\omega)}
\right) \right] .
\label{eq:dosfull}
\end{eqnarray}
Note that in the absence of impurities ($\Gamma(\omega)=0$),
\begin{equation}
N(\omega) \Bigr|_{\Gamma(\omega)=0} = \frac{|\omega|}{\pi v_{f} v_{2}}
\label{eq:dosGamma0}
\end{equation}
while in the presence of impurities, there is a finite density of quasiparticle
states down to zero energy \cite{gor85}:
\begin{equation}
N(0) = \frac{2}{\pi^{2} v_{f} v_{2}} \Gamma_{0} \ln\frac{p_{0}}{\Gamma_{0}}
\label{eq:dos0}
\end{equation}
where $\Gamma_{0}\equiv\Gamma(\omega\rightarrow0)$.  These
{\em impurity-induced} quasiparticles are responsible for the intriguing
low temperature transport properties that we shall consider in the sections
that follow.

\section{Microwave Electrical Conductivity}
\label{sec:elec}
Electrical conductivity can be calculated by means of the Kubo formula
\cite{mah90}
\begin{equation}
\sigma(\Omega,T) = - \frac{\mbox{Im}\, \Pi_{ret}(\Omega)}{\Omega}
\label{eq:econddef}
\end{equation}
where $\Pi_{ret}(\Omega) = \Pi(i\Omega\rightarrow\Omega+i\delta)$
and $\Pi(i\Omega)$ is the current-current correlation function (or 
polarization function) in the Matsubara finite temperature formalism:
\begin{equation}
\tensor{\Pi}(i\Omega) = - \int_{0}^{\beta} d\tau\, e^{i\Omega\tau}
\langle T_{\tau} {\bf j}^{\dagger}(\tau) {\bf j}(0) \rangle .
\label{eq:matpolfunc}
\end{equation}
Thus, our first step is to derive an expression for the electrical current
operator.  Then by evaluating its correlation function, we obtain the
electrical conductivity.

\subsection{Electrical Current}
\label{ssec:eleccurrent}
For a system of interacting electrons, the Hamiltonian is given by
\begin{eqnarray}
H = && \int d{\bf x} \psi_{\alpha}^{\dagger}({\bf x})
\left(\frac{-\nabla^{2}}{2m^{*}}\right) \psi_{\alpha}({\bf x}) \nonumber\\
&& + \frac{1}{2} \int d{\bf x}d{\bf y} \psi_{\alpha}^{\dagger}({\bf x})
\psi_{\beta}^{\dagger}({\bf y}) V({\bf x} - {\bf y})
\psi_{\beta}({\bf y}) \psi_{\alpha}({\bf x})
\label{eq:elecham}
\end{eqnarray}
where $\psi_{\alpha}({\bf x})$ annihilates an electron of spin $\alpha$ at
position ${\bf x}$ and $V({\bf x} - {\bf y})$ is the electron-electron
interaction potential.  In the presence of a vector potential
${\bf A}({\bf x})$, the Hamiltonian must be invariant under a local
gauge transformation.  While the second term is gauge invariant as written,
the first must be modified by making the standard replacement
$-i\nabla \rightarrow -i\nabla + e{\bf A}$.  Thus, the Hamiltonian becomes
a functional of the vector potential and takes the form
\begin{eqnarray}
H[{\bf A} && ({\bf x})] = \int d{\bf x} \psi_{\alpha}^{\dagger}({\bf x})
\left(\frac{(-i\nabla + e{\bf A})^{2}}{2m^{*}}\right) \psi_{\alpha}({\bf x})
\nonumber\\
&& + \frac{1}{2} \int d{\bf x}d{\bf y} \psi_{\alpha}^{\dagger}({\bf x})
\psi_{\beta}^{\dagger}({\bf y}) V({\bf x} - {\bf y})
\psi_{\beta}({\bf y}) \psi_{\alpha}({\bf x}) .
\label{eq:elechamA}
\end{eqnarray}
Note that only the kinetic term couples to the vector potential.
Taking the functional derivative with respect to ${\bf A}({\bf x})$ we
obtain an expression for the electrical current:
\begin{eqnarray}
{\bf j}^{e}({\bf x}) &=& - \mbox{Re} \,
\frac{\delta H}{\delta {\bf A}({\bf x})} \nonumber\\
&=& \frac{-e}{2im^{*}} \left( \psi^{\dagger}_{\alpha} \nabla \psi_{\alpha}
- \nabla \psi^{\dagger}_{\alpha} \psi_{\alpha} \right) .
\label{eq:elecjx}
\end{eqnarray}
Then taking the space-time Fourier transform yields
\begin{equation}
{\bf j}^{e}({\bf q},\Omega) = - \frac{e}{m^{*}} \sum_{k,\omega}
\left( {\bf k} + \frac{{\bf q}}{2} \right) c^{\dagger}_{k \alpha}
c_{k+q \alpha}
\label{eq:elecjq}
\end{equation}
and in the limit that ${\bf q} \rightarrow 0$ we obtain
\begin{equation}
{\bf j}^{e}(0,\Omega) = -e \sum_{k,\omega} {\bf v}_{f}
\Psi^{\dagger}_{k} \tilde{\openone} \Psi_{k+q}
\label{eq:elecjqnambu}
\end{equation}
where ${\bf v}_{f} \equiv \partial\epsilon_{k}%
/ \partial {\bf k}={\bf k}/m^{*}$
and we have expressed the final result in terms of $2 \times 2$
Nambu matrix notation.  Note that in all cases, momentum indices on
field operators denote both momentum and frequency.
Although this result is well known, we have derived it here in order to
provide a basis for comparison with the thermal current and spin current
to be derived later.

\subsection{Bare Bubble (Electrical)}
Given the current, we proceed to calculate
the corresponding current-current correlation function.
The correlation function for the electrical current can be expressed
diagramatically as a fermionic bubble with fully dressed propagators
and a fully dressed vertex in which each vertex contributes a coupling
parameter $e$, a velocity ${\bf v}_{f}$, and a $2 \times 2$ Nambu formalism unit
matrix $\tilde{\openone}$.  Assuming that the impurity scattering
potential is isotropic in k-space, the corrections to the bare vertex
vanish.  Thus, in this approximation, the conductivity can be obtained
from the calculation of a bubble with dressed propagators (i.e.\ Green's
functions with self-energy included) but bare vertices (no interaction
between the two propagators).  Such a diagram will be referred to as a
{\em bare bubble}.

The calculation of the bare bubble polarization function is of the same
basic form for the electrical, thermal, and spin conductivities.
Thus, to avoid repeating the same
derivation several times, a generalized polarization function
$\Pi^{g\ell\alpha}_{ret}(\Omega)$ (applicable to all three cases)
which depends on a coupling parameter $g$, a velocity ${\bf v}_{\ell}$,
and a Nambu matrix $\tilde{\tau}_{\alpha}$ has been calculated in
Appendix \ref{app:barebubble}.  Applying the general
result (\ref{eq:imretpolbare}) to the case of interest ($g=e$,
${\bf v}_{\ell} = {\bf v}_{f}$, $\tilde{\tau}_{\alpha} = \tilde{\openone}$)
we find that
\begin{eqnarray}
\sigma(\Omega,T) = \frac{e^{2}}{\pi^{2}} \frac{v_{f}}{v_{2}} &&
\int \frac{d^{2}p}{2\pi} \int_{-\infty}^{\infty} d\omega
\frac{n_{F}(\omega) - n_{F}(\omega+\Omega)}{\Omega} \nonumber \\
&& \times \mbox{Tr} \left[ \tilde{G}''_{ret}({\bf p},\omega)
\tilde{G}''_{ret}({\bf p},\omega+\Omega) \right]
\label{eq:eleccondbare1}
\end{eqnarray}
where
\begin{eqnarray}
\tilde{G}_{ret} && ({\bf p},\omega) = \frac{1}{(\omega - \Sigma_{ret}(\omega))^{2}
- p^{2}} \nonumber \\
&& \times \left( \begin{array}{cc}
\omega - \Sigma_{ret}(\omega) + p_{1} & p_{2} \\
p_{2} & \omega - \Sigma_{ret}(\omega) - p_{1}
\end{array} \right) .
\label{eq:ebGret}
\end{eqnarray}
In the universal limit ($\Omega\rightarrow0$, $T\rightarrow0$),
\begin{equation}
\frac{n_{F}(\omega) - n_{F}(\omega+\Omega)}{\Omega}
\rightarrow - \frac{\partial n_{F}}{\partial \omega}
\rightarrow \delta(\omega) .
\label{eq:diff2deriv2delta}
\end{equation}
Hence, evaluating the rest of the integrand for $\omega\rightarrow0$,
noting that $\Sigma_{ret}(0)=-i\Gamma_{0}$, and
integrating over momentum, we obtain the universal limit bare bubble
electrical conductivity:
\begin{equation}
\sigma_{0} = \frac{e^{2}}{\pi^{2}} \frac{v_{f}}{v_{2}} .
\label{eq:eleccondbare}
\end{equation}
This is the universal conductivity obtained in Ref.\ \onlinecite{lee93}.
Finite temperature corrections can be obtained
via a Sommerfeld expansion (for $T \ll \Gamma(\omega)$)
and have been calculated by Hirschfeld {\it et al.\cite{hir94}}
and Graf {\it et al.\cite{gra96}}

\subsection{Vertex Corrections (Electrical)}
\label{ssec:elecvert}
The bare bubble conductivity derived above was calculated in the
approximation that vertex corrections could be safely neglected.
It turns out (as we shall see later) that this approximation is
justified if the impurity scattering potential is isotropic in
k-space ($V_{kk'}=V=\mbox{const}$).  However, for a general
scattering potential, corrections to the bare vertex can make
a significant contribution and must be included in the calculation.
To this end, we shall consider the contribution of the
{\em ladder corrections} to the bare vertex (see Fig.\
\ref{fig:vertcorrbub} of Appendix \ref{app:vertexcorrections}).
Once again, since the electrical calculation is of the
same form as that for the thermal and spin conductivity, a
generalized polarization function including vertex corrections
(which can be applied to all three cases) has been calculated in
Appendix \ref{app:vertexcorrections}.  For the electrical conductivity,
the polarization function consists of a single bubble (with a dressed
vertex) where each vertex contributes a coupling parameter $e$, a velocity
${\bf v}_{f}$, and a Nambu matrix $\tilde{\openone}$.  Plugging these
parameters into the generalized polarization function
(\ref{eq:imretpolvert}) and using the electrical Kubo formula
(\ref{eq:econddef}) we find that
\begin{eqnarray}
\sigma(\Omega,T) = && \frac{e^{2}}{2\pi^{2}} \frac{v_{f}}{v_{2}}
\int_{-\infty}^{\infty} \!\!\!\!\!\! d\omega\ \frac{n_{F}(\omega)
- n_{F}(\omega+\Omega)}{\Omega} \nonumber \\
&& \times \mbox{Re} \left[ J^{(0)}_{2}(\omega,\Omega)
- J^{(0)}_{1}(\omega,\Omega) \right]
\label{eq:eleccondvert1}
\end{eqnarray}
where $J^{\alpha}_{1}$ and $J^{\alpha}_{2}$ are defined in Appendix
\ref{app:vertexcorrections}.
In the universal limit ($\Omega\rightarrow0$, $T\rightarrow0$),
we can make use of (\ref{eq:diff2deriv2delta}) to find that
\begin{eqnarray}
\sigma_{0} = \frac{e^{2}}{2\pi^{2}} \frac{v_{f}}{v_{2}} \mbox{Re}
&& \left[ \frac{I^{(0)}_{2}}{1 - \gamma^{(0)}_{A2} I^{(0)}_{2}
\left( 1 + \frac{\gamma^{(0)}_{B2}}{\gamma^{(0)}_{A2}}
\frac{\gamma^{(0)}_{B2} I^{(1)}_{2}}{1 - \gamma^{(0)}_{A2} I^{(1)}_{2}}
\right) } \right. \nonumber \\
&& \left. - \frac{I^{(0)}_{1}}{1 - \gamma^{(0)}_{A1} I^{(0)}_{1}
\left( 1 + \frac{\gamma^{(0)}_{B1}}{\gamma^{(0)}_{A1}}
\frac{\gamma^{(0)}_{B1} I^{(1)}_{1}}{1 - \gamma^{(0)}_{A1} I^{(1)}_{1}}
\right) } \right] \nonumber \\ &&
\label{eq:eleccondvert2}
\end{eqnarray}
where all functions are evaluated for \mbox{$\Omega,\omega\rightarrow0$}.
In these limits, the constituent
functions defined in Appendix \ref{app:vertexcorrections} take the form:
\begin{equation}
F'(0) = \frac{F(i\Gamma_{0})}{4\pi v_{f} v_{2}}
= i\frac{\Gamma_{0}}{2\pi v_{f} v_{2}} \ln \frac{p_{0}}{\Gamma_{0}}
= i \frac{\pi}{4} N(0)
\label{eq:F'0}
\end{equation}
\begin{equation}
T^{a}_{n}(0) = \left( \frac{\underline{V}}{1 + (\frac{\pi}{4}N(0))^{2}
\underline{V}^{2}} \right)_{n1} \equiv A_{n}
\label{eq:Tmata0}
\end{equation}
\begin{equation}
T^{b}_{n}(0) = i \left( \frac{-\frac{\pi}{4} N(0) \underline{V}^{2}}
{1 + (\frac{\pi}{4}N(0))^{2} \underline{V}^{2}} \right)_{n1} \equiv iB_{n}
\label{eq:Tmatb0}
\end{equation}
\begin{mathletters}
\label{eq:gammas0}
\begin{eqnarray}
\gamma^{(0)}_{A1} &=& \frac{n_{i}}{4\pi v_{f} v_{2}}
\left( A_{n}^{2} - B_{n}^{2} \right) \bigl[ |_{n=1} - |_{n=3} \bigr] \\
\gamma^{(0)}_{A2} &=& \frac{n_{i}}{4\pi v_{f} v_{2}}
\left( A_{n}^{2} + B_{n}^{2} \right) \bigl[ |_{n=1} - |_{n=3} \bigr] \\
\gamma^{(0)}_{B1} &=& i \frac{n_{i}}{2\pi v_{f} v_{2}}
A_{n}B_{n} \bigl[ |_{n=1} - |_{n=3} \bigr] \\
\gamma^{(0)}_{B2} &=& 0
\end{eqnarray}
\end{mathletters}
\begin{equation}
I^{(0)}_{1}(0,0) = \left. \frac{dF(z)}{dz} \right|_{i\Gamma_{0}}
= 2 \ln\frac{p_{0}}{\Gamma_{0}} - 2
\label{eq:I(0)1}
\end{equation}
\begin{equation}
I^{(0)}_{2}(0,0) = \frac{\mbox{Im}\,F(i\Gamma_{0})}{\Gamma_{0}}
= 2 \ln\frac{p_{0}}{\Gamma_{0}}
\label{eq:I(0)2}
\end{equation}
\begin{equation}
I^{(1)}_{1}(0,0) = \frac{1}{2} \left(
\left. \frac{dF(z)}{dz} \right|_{i\Gamma_{0}}
- \frac{F(i\Gamma_{0})}{i\Gamma_{0}} \right) = -1
\label{eq:I(1)1}
\end{equation}
\begin{equation}
I^{(1)}_{2}(0,0) = \lim_{\omega \rightarrow 0}
\frac{\mbox{Im} \left[ (\omega - i\Gamma_{0}) F(\omega + i\Gamma_{0})
\right] }{2 \Gamma_{0} \omega} = 1
\label{eq:I(1)2}
\end{equation}
Thus, including vertex corrections,
the universal limit electrical conductivity takes the form
\begin{equation}
\sigma_{0} = \frac{e^{2}}{\pi^{2}} \frac{v_{f}}{v_{2}} \beta_{vc}
\label{eq:eleccondvert}
\end{equation}
\begin{eqnarray}
&& \beta_{vc} = \nonumber \\
&& \frac{1 + 2 \left( \gamma^{(0)}_{A2} - \gamma^{(0)}_{A1}
+\frac{\left. \gamma^{(0)}_{B1} \right.^{2}}{1-\gamma^{(0)}_{A1}} \right)
\ln\frac{p_{0}}{\Gamma_{0}} \left( \ln\frac{p_{0}}{\Gamma_{0}} - 1 \right)}
{\left( 1 - 2\gamma^{(0)}_{A2}
\ln\frac{p_{0}}{\Gamma_{0}} \right) \left( 1 - 2\left( \gamma^{(0)}_{A1} -
\frac{\left. \gamma^{(0)}_{B1} \right.^{2}}{1-\gamma^{(0)}_{A1}} \right)
\left( \ln\frac{p_{0}}{\Gamma_{0}} - 1 \right) \right) } \nonumber \\ &&
\label{eq:betaVC}
\end{eqnarray}
where $\beta_{vc}$ is the scattering dependent vertex correction to the universal
bare bubble result.
Note that since the $\gamma$'s all depend on the difference between the 
intra-node T-matrix (n=1) and the opposite-node T-matrix (n=3),
$\beta_{vc}\rightarrow1$ if the two scattering potentials are the same.
Hence for an isotropic scattering potential,
the bare bubble result (\ref{eq:eleccondbare}) is recovered.
However, in general we presume that the scattering potential will fall off for
large ${\bf k}$ and the potential for intra-node scattering will be larger than
that for opposite-node scattering.  If so, the $\gamma$'s will be non-zero and
the universal limit conductivity will deviate from its bare bubble value.  This
correction to the conductivity due to differences between intra-node
(forward) scattering and opposite-node (back) scattering is the
node-discrete equivalent of the famous $1 - \cos\theta$ factor obtained from 
vertex corrections in the conductivity calculation for a simple metal.  As in
the metallic case, the phenomenon at work is the fact that forward and back
scattering can have different effects on the progress of a charge carrier.
As a result, anisotropy in the scattering potential can renormalize the
conductivity.

In general, the evaluation of $\beta_{vc}$ requires a numerical calculation
since $\Gamma_{0}$ must be obtained self-consistently as a function of
impurity density and scattering potential.  Such calculations (presented
in Appendix \ref{app:numerical}) indicate that for anisotropic scattering,
the electrical vertex correction can be significant even to zeroth order
in the impurity density.

For the case of Born scattering, $\beta_{vc}$
reduces to a more simple and illustrative form.  In the Born limit
(small V),
\begin{equation}
T_{n}^{a}(0) = V_{n1} \;\;\;\;\;\;\;\; T_{n}^{b}(0) = 0
\label{eq:bornTa}
\end{equation}
\begin{equation}
\gamma^{(0)}_{A1} = \gamma^{(0)}_{A2} = \frac{n_{i}}{4\pi v_{f} v_{2}}
(V_{1}^{2} - V_{3}^{2}) \;\;\;\;\;\;\;\; \gamma^{(0)}_{B1} = 0
\label{eq:borngammaA}
\end{equation}
and the zero-frequency scattering rate takes the form
\begin{eqnarray}
\Gamma_{0} &=& p_{0} \exp \left( -\frac{2 \pi v_{f} v_{2}}{n_{i}
(V_{1}^{2} + 2 V_{2}^{2} + V_{3}^{2})} \right) \nonumber \\
&=& \frac{\pi}{4} n_{i} (V_{1}^{2} + 2 V_{2}^{2} + V_{3}^{2}) N(0)
\label{eq:bornGamma0}
\end{eqnarray}
where (as defined in Sec. \ref{sec:dos}) $V_{1}$,
$V_{2}$, and $V_{3}$ correspond respectively to intra-node, adjacent-node,
and opposite-node scattering.  Noting that
$\ln\frac{p_{0}}{\Gamma_{0}} \sim \frac{1}{n_{i}} \gg 1$ and defining
\begin{equation}
\Gamma_{1} \equiv 2 \gamma^{(0)}_{A1} \Gamma_{0} \ln \frac{p_{0}}{\Gamma_{0}}
= \frac{\pi}{4} n_{i} (V_{1}^{2} - V_{3}^{2}) N(0)
\label{eq:bornGamma1}
\end{equation}
and a transport scattering rate
\begin{equation}
\Gamma_{tr} \equiv \Gamma_{0} - \Gamma_{1} = \frac{\pi}{4} n_{i}
(2 V_{2}^{2} + 2 V_{3}^{2}) N(0)
\label{eq:bornGammatr}
\end{equation}
the vertex correction factor (\ref{eq:betaVC}) reduces to
\begin{equation}
\beta_{vc} = \left( \frac{\Gamma_{0}}{\Gamma_{tr}} \right)^{2}
= \left( \frac{V_{1}^{2} + 2 V_{2}^{2} + V_{3}^{2}}{2 V_{2}^{2} + 2 V_{3}^{2}}
\right)^{2} .
\label{eq:bornbetaVC}
\end{equation}
Note that the vertex correction depends on the scattering potential but is
independent of the density of impurities.  In this simple limit it is clear that
if intra-node scattering is stronger than opposite-node scattering (as we expect),
$\Gamma_{0}$ will exceed $\Gamma_{tr}$ and the universal limit electrical
conductivity will be enhanced beyond the bare bubble result.

\subsection{Fermi Liquid Corrections (Electrical)}
\label{ssec:elecfermi}
To this point, our calculation of electrical conductivity has neglected the
effects of the underlying Fermi liquid interaction between electrons.
In Sec.\ \ref{ssec:eleccurrent}, an explicit
expression was derived for the electrical current in the absence of Fermi
liquid interactions.  In essence, it has the form
\begin{equation}
{\bf j}^{e}_{0} = -e \sum_{k\alpha} {\bf v}_{fk} \,\delta n_{k\alpha}
\label{eq:bareeleccurrent}
\end{equation}
where ${\bf v}_{fk}$ is the Fermi velocity at ${\bf k}$ and
$\delta n_{k\alpha}$ is the deviation of the electron distribution from
equilibrium.  The Fermi liquid renormalization of such a current has been
derived in Appendix \ref{app:fermiliquid}.  Plugging into the general result
(\ref{eq:dresscurrent}) we see that in the presence of Fermi liquid
interactions, the electrical current is given by
\begin{eqnarray}
{\bf j}^{e} = && {\bf j}^{e}_{0} - e \sum_{k'\alpha'} \delta n_{k'\alpha'}
\sum_{k} {\bf v}_{fk}\, f^{s}_{kk'}
\Biggl[ \frac{\Delta_{k}^{2}}
{2 ( \epsilon_{k}^{2} + \Delta_{k}^{2} )^{3/2}} \nonumber \\
&& + \left( \frac{\epsilon_{k}^{2}}{E_{k}^{2}}
\frac{\Gamma_{0}/\pi}{E_{k}^{2} + \Gamma_{0}^{2}}
- \frac{\Delta_{k}^{2}}{\pi E_{k}^{3}} \arctan \left( \frac{\Gamma_{0}}{E_{k}}
\right) \right) \Biggr]
\label{eq:dresseleccurrent1}
\end{eqnarray}
where $f^{s}_{kk'} = f^{\uparrow\uparrow}_{kk'} + f^{\downarrow\uparrow}_{kk'}$.
The first term in brackets is peaked at $\epsilon_{k}=0$ and is therefore a
Fermi surface term (smeared over the extent of the gap).
If we assume a circular Fermi surface, then
replacing the k-sum by an integral in circular coordinates
$(k,\theta)$, presuming that $\Delta_{k}=\Delta(\theta) \ll E_{F}$,
and expanding the Landau function in 2D harmonics
\begin{equation}
f^{s}(\theta - \theta') = \frac{1}{\nu(0)} \sum_{\ell=0}^{\infty} F^{s}_{\ell}
\cos \left( \ell ( \theta - \theta') \right)
\label{eq:harmonics}
\end{equation}
this first term takes the form
\begin{equation}
{\bf j}^{e}_{1} = {\bf j}^{e}_{0} \frac{F^{s}_{1}}{2}
\label{eq:jeterm1}
\end{equation}
where $F^{s}_{1}$ is the $\ell=1$ spin-symmetric Landau parameter and
$\nu(0)$ is the single spin normal state density of states at the Fermi surface.
The second term in brackets is peaked at $E_{k}=0$ and is therefore a node term
(contributing primarily at the gap nodes).  Replacing the k-sum
by an integral about each of the nodes and noting by symmetry that
\begin{equation}
\sum_{j=1}^{4} {\bf v}_{f}^{j} f^{s}_{jj'} = {\bf v}_{f}^{j'}
\left( f^{s}_{11} - f^{s}_{31} \right)
\label{eq:landaunodesum}
\end{equation}
we find that
\begin{equation}
{\bf j}^{e}_{2} = - {\bf j}^{e}_{0} \frac{\Gamma_{0}
\left( f^{s}_{11} - f^{s}_{31} \right)}{4 \pi^{2} v_{f} v_{2}}
\label{eq:jeterm2}
\end{equation}
where $f^{s}_{11}$ and $f^{s}_{31}$ are respectively the intra-node and
opposite-node spin-symmetric Fermi liquid interaction energies.
In the small impurity density limit, the second (node) term can be neglected
with respect to the first (Fermi surface) term.
Hence, the renormalized electrical current is given by
\begin{equation}
{\bf j}^{e} = {\bf j}^{e}_{0} \alpha_{fl}^{s}
\label{eq:dresseleccurrent}
\end{equation}
where (for a circular Fermi surface)
\begin{equation}
\alpha_{fl}^{s} = 1 + \frac{F^{s}_{1}}{2}
- \frac{\Gamma_{0} \left( f^{s}_{11} - f^{s}_{31} \right)}
{4 \pi^{2} v_{f} v_{2}}
\approx 1 + \frac{F^{s}_{1}}{2}
\label{eq:alphaeleccircular}
\end{equation}
and the superscript `s' denotes that this is the spin-symmetric
current renormalization factor.
The simple form of this expression is due to our assumption of a circular
Fermi surface.  For a more general Fermi surface, additional harmonics
of the Landau function would have been generated.  Thus, in practice,
this factor should be treated as a parameter to be determined by experiment.
Note that the renormalization
applies to both the quasiparticle (normal) current and the supercurrent.
However, since the supercurrent does not contribute to the real part of
the AC conductivity, we are concerned only with the normal current.

The basic physics of this renormalization is as follows.
Upon the application of an electric field,
quasiparticles are perturbed to form a normal current.  In the presence of
the excited quasiparticles, the electron dispersion is modified due to
the Fermi liquid interaction.  The modified dispersion yields a modified
equilibrium distribution which means that the deviation from equilibrium is
also modified and the current is renormalized.  Note that the dominant term
in the renormalization factor is a Fermi surface term resulting from the
modification of the equilibrium condensate distribution in the
presence of perturbed quasiparticles.

So far, we have discussed only how Fermi liquid interactions
renormalize the electrical current density operator.  Yet our goal is to
determine the manner in which such interactions modify the
electrical conductivity.  In general, such modifications can be more
complicated than merely renormalizing the constituent currents.
However, as discussed in Appendix \ref{app:fermiliquid}, current
renormalization is the dominant effect in the $T\rightarrow0$ limit with
which we are concerned.  Therefore, since electrical conductivity
is proportional to the current-current correlation function, two powers of
our current renormalization factor appear in the conductivity.  Hence, including
both vertex corrections and Fermi liquid corrections, the electrical
conductivity in the universal limit takes the form
\begin{equation}
\sigma_{0} = \frac{e^{2}}{\pi^{2}} \frac{v_{f}}{v_{2}} \beta_{vc}
\left. \alpha_{fl}^{s} \right.^{2} .
\label{eq:eleccondfermi}
\end{equation}

\subsection{Superfluid Density}
As a check on the accuracy of our conductivity calculations, it is useful to
make a brief digression and use our results to calculate an experimentally
distinct quantity, the superfluid density (without impurities), $\rho^{s}(T)$.
By definition (see Ref.\ \onlinecite{lee97}),
\begin{equation}
\rho^{s}(T) \equiv \rho^{s}(T=0) - \rho^{n}(T)
\label{eq:rhosrhon}
\end{equation}
where $\rho^{n}(T)$ is the normal fluid density.  Hence, to obtain
the temperature dependence of the superfluid density, it suffices to
calculate the normal fluid density.
While the conductivity is related to the imaginary part of the polarization
function, the normal fluid density is proportional to the real part via
\begin{equation}
\frac{\rho^{n}(T)}{m} = - \frac{ \mbox{Re}\, \Pi_{ret}(\Omega=0)}{e^{2}} .
\label{eq:rhodef}
\end{equation}
Obtaining $\Pi_{ret}$ from the generalized result (\ref{eq:retpolvert})
in Appendix \ref{app:vertexcorrections}, setting $\Omega=0$, taking the
no impurities limit ($\Gamma(\omega)\rightarrow0$), and plugging into
(\ref{eq:rhodef}) we find that
\begin{equation}
\frac{\rho^{n}(T)}{m} = \frac{1}{\pi^{2}} \frac{v_{f}}{v_{2}}
\int_{-\infty}^{\infty} \!\!\!\!\!\! d\omega\ n_{F}(\omega)\
\mbox{Im} \left[ I^{(0)}_{1}(\omega,0) \right]
\label{eq:rho1}
\end{equation}
where
\begin{eqnarray}
\mbox{Im} \left[ I^{(0)}_{1}(\omega,0) \right]
= && \pi \mbox{sgn}(\omega) \left[ \theta(\omega+p_{0}) - \theta(\omega-p_{0})
\right] \nonumber \\
&& + \pi p_{0} \left[ \delta(\omega+p_{0}) - \delta(\omega-p_{0}) \right] .
\label{eq:ImI1}
\end{eqnarray}
Performing the frequency integration yields that the normal fluid density
neglecting Fermi liquid corrections is given by
\begin{equation}
\frac{\rho^{n}(T)}{m} = \frac{2 \ln 2}{\pi} \frac{v_{f}}{v_{2}} k_{B} T
\label{eq:rhonofermi}
\end{equation}
which is precisely the result obtained in Ref.\ \onlinecite{lee97} through
an entirely different procedure.
To include Fermi liquid corrections in the $T\rightarrow0$ limit, we
note (via Appendix \ref{app:fermiliquid}) that the primary effect
of Fermi liquid interactions is the renormalization of the current density
operator.  Thus, since the normal fluid density is proportional to the
current-current correlation function, we
need only multiply by two factors of the current renormalization
(obtained in the previous section) to find that
\begin{equation}
\frac{\rho^{n}(T)}{m} = \frac{2 \ln 2}{\pi} \frac{v_{f}}{v_{2}}
\left. \alpha_{fl}^{s} \right.^{2} k_{B} T
\label{eq:rhofermi}
\end{equation}
which agrees with the results of Refs.\ \onlinecite{mil98,xu95,leg65}.
Given this correspondence with prior work, we can be reassured of the
accuracy of our calculations.

\section{Thermal Conductivity}
\label{sec:therm}
Analogous to the case of electrical conductivity, thermal conductivity can
be calculated by means of a thermal Kubo formula \cite{mah90}:
\begin{equation}
\frac{\kappa(\Omega,T)}{T} = - \frac{1}{T^{2}} \frac{\mbox{Im}\,
\Pi^{\kappa}_{ret}(\Omega)}{\Omega}
\label{eq:tconddef}
\end{equation}
where $\Pi^{\kappa}_{ret}(\Omega)=\Pi^{\kappa}(i\Omega \rightarrow \Omega
+ i\delta)$ and $\Pi^{\kappa}(i\Omega)$ is the finite temperature
current-current correlation function (or polarization function).
In this case, the appropriate current for the correlation function
is the thermal current derived below.

\subsection{Thermal Current}
To derive an expression for the heat current in an anisotropic superconductor,
we can follow the s-wave derivation of Ambegaokar and Griffin \cite{amb65}
and generalize to the case of an anisotropic gap.  As in (\ref{eq:elecham}),
the Hamiltonian takes the form
\begin{eqnarray}
H = && \int d{\bf x} \psi_{\alpha}^{\dagger}({\bf x})
\left(\frac{-\nabla^{2}}{2m^{*}}\right) \psi_{\alpha}({\bf x}) \nonumber\\
&& + \frac{1}{2} \int d{\bf x}d{\bf y} \psi_{\alpha}^{\dagger}({\bf x})
\psi_{\beta}^{\dagger}({\bf y}) V({\bf x} - {\bf y})
\psi_{\beta}({\bf y}) \psi_{\alpha}({\bf x})
\label{eq:ham}
\end{eqnarray}
Given the Hamiltonian, it is straightforward to
obtain the equations of motion for the field operators
\begin{eqnarray}
i\dot{\psi}_{\alpha}
&& = \left[\psi_{\alpha},H\right] \nonumber\\
&& = \left( \frac{-\nabla^{2}}{2m^{*}} + \int d{\bf r} V({\bf x} - {\bf r})
\psi_{\gamma}^{\dagger}({\bf r}) \psi_{\gamma}({\bf r}) \right)
\psi_{\alpha}
\label{eq:eom}
\end{eqnarray}
and to define a Hamiltonian density
\begin{eqnarray}
h({\bf x}) = && \frac{1}{2m^{*}} \nabla\psi_{\alpha}^{\dagger}({\bf x})
\cdot \nabla\psi_{\alpha}({\bf x}) \nonumber\\
&& + \frac{1}{2} \int d{\bf y} V({\bf x} - {\bf y})
\psi_{\alpha}^{\dagger}({\bf x}) \psi_{\beta}^{\dagger}({\bf y})
\psi_{\beta}({\bf y}) \psi_{\alpha}({\bf x})
\label{eq:hamden}
\end{eqnarray}
If all energies are measured with respect to the chemical potential,
this Hamiltonian density is the heat density.  Hence, the operator
${\bf j}^{Q}({\bf x})$ that satisfies the continuity equation
\begin{equation}
\dot{h}({\bf x}) + \nabla \cdot {\bf j}^{Q}({\bf x}) = 0
\label{eq:continuity}
\end{equation}
can be interpreted as the heat current.  Taking the time derivative of
(\ref{eq:hamden}) and using the equations of motion (\ref{eq:eom})
we find that
\begin{eqnarray}
\dot{h} = && \nabla\cdot (\dot{\psi}_{x\alpha}^{\dagger}
\nabla\psi_{x\alpha} + \nabla\psi_{x\alpha}^{\dagger}
\dot{\psi}_{x\alpha}) - \frac{1}{2}\int d{\bf y}
V({\bf y} - {\bf x}) \nonumber\\
&& \times  \left[ (\dot{\psi}_{x\alpha}^{\dagger} \psi_{y\beta}^{\dagger}
\psi_{y\beta} \psi_{x\alpha}
+ \psi_{x\alpha}^{\dagger} \psi_{y\beta}^{\dagger}
\psi_{y\beta} \dot{\psi}_{x\alpha}) \right. \nonumber\\
&& \left. - (\psi_{x\alpha}^{\dagger} \dot{\psi}_{y\beta}^{\dagger}
\psi_{y\beta} \psi_{x\alpha}
+ \psi_{x\alpha}^{\dagger} \psi_{y\beta}^{\dagger}
\dot{\psi}_{y\beta} \psi_{x\alpha}) \right]
\label{eq:hdot}
\end{eqnarray}
where the compact notation $\psi_{x\alpha} \equiv \psi_{\alpha}({\bf x})$
has been used for the sake of brevity.  Defining
${\bf j}^{Q} = {\bf j}^{Q}_{1} + {\bf j}^{Q}_{2}$ we can use
(\ref{eq:continuity}) to write
\begin{equation}
{\bf j}^{Q}_{1}({\bf x}) = -\frac{1}{2m^{*}} \left( \dot{\psi}_{x\alpha}^{\dagger}
\nabla\psi_{x\alpha} + \nabla\psi_{x\alpha}^{\dagger}
\dot{\psi}_{x\alpha} \right)
\label{eq:j1x}
\end{equation}
and
\begin{eqnarray}
\nabla\cdot{\bf j}^{Q}_{2}({\bf x}) &=& \frac{1}{2}\int d{\bf y}
V({\bf y} - {\bf x}) \nonumber\\
&\times& \left[ (\dot{\psi}_{x\alpha}^{\dagger} \psi_{y\beta}^{\dagger}
\psi_{y\beta} \psi_{x\alpha}
+ \psi_{x\alpha}^{\dagger} \psi_{y\beta}^{\dagger}
\psi_{y\beta} \dot{\psi}_{x\alpha}) \right. \nonumber\\
&-& \left. (\psi_{x\alpha}^{\dagger} \dot{\psi}_{y\beta}^{\dagger}
\psi_{y\beta} \psi_{x\alpha}
+ \psi_{x\alpha}^{\dagger} \psi_{y\beta}^{\dagger}
\dot{\psi}_{y\beta} \psi_{x\alpha}) \right] .
\label{eq:j2x}
\end{eqnarray}

Taking the space-time Fourier transform of (\ref{eq:j1x}) in the limit
that ${\bf q} \rightarrow 0$ we obtain
\begin{eqnarray}
{\bf j}^{Q}_{1}(0,\Omega) &=&
\sum_{k,\omega}\left(\omega + \frac{\Omega}{2}\right){\bf v}_{f}
c_{k\alpha}^{\dagger} c_{k+q\alpha} \nonumber\\
&=& \sum_{k,\omega}\left(\omega + \frac{\Omega}{2}\right){\bf v}_{f}
\Psi_{k}^{\dagger} \tilde{\tau}_{3} \Psi_{k+q}
\label{eq:j1q}
\end{eqnarray}
where ${\bf v}_{f} \equiv \partial\epsilon_{k}%
/ \partial {\bf k}={\bf k}/m^{*}$,
$c_{k\alpha}$ is the space-time Fourier transform of $\psi_{\alpha}({\bf x})$,
and the second line is written in terms of the 2x2 Nambu matrix
notation introduced in Sec. \ref{sec:dos}.  Please note that in our compact notation,
momentum indices on field operators always represent both momentum and
frequency  (i.\ e.\ $\Psi_{k} \equiv \Psi({\bf k},\omega)$ and
$\Psi_{k+q} \equiv \Psi({\bf k}+{\bf q},\omega + \Omega)$).

Similarly, taking the space-time Fourier transform of (\ref{eq:j2x})
generates four terms such that
\begin{equation}
i{\bf q} \cdot {\bf j}^{Q}_{2}({\bf q},\Omega)
= X_{1} + X_{2} - Y_{1} - Y_{2} .
\label{eq:j2qsumof4}
\end{equation}
The first such term, $X_{1}$ , is given by
\begin{eqnarray}
X_{1} = && \sum_{k's , \omega's}
\omega_{1} V_{k_{5}} c_{k_{1}\alpha}^{\dagger} c_{k_{2}\beta}^{\dagger}
c_{k_{3}\beta} c_{k_{4}\alpha} \nonumber\\
&& \times \delta_{k_{4}-k_{1}-k_{5}-q} \delta_{k_{3}-k_{2}+k_{5}}
\delta_{\omega_{1}+\omega_{2}-\omega_{3}-\omega_{4}+\Omega} .
\label{eq:X1init}
\end{eqnarray}
Taking the mean field approximation, retaining only the terms for which
the average values are over $({\bf k}\uparrow , -{\bf k}\downarrow)$ pairs,
and using the fact that
$\langle c_{k \uparrow}^{\dagger} c_{-k \downarrow}^{\dagger} \rangle$
is an even function of $\omega$ , this becomes
\begin{equation}
X_{1} = -i\sum_{k,\omega} (\omega - \Omega) \Delta_{k}^{\dagger}
c_{k-q \uparrow}^{\dagger} c_{-k \downarrow}^{\dagger}
\label{eq:X1meanred}
\end{equation}
where
\begin{equation}
\Delta_{k} \equiv -\sum_{k' , \omega'} V_{k-k'}
\left\langle c_{k \uparrow}^{\dagger} c_{-k \downarrow}^{\dagger}
\right\rangle .
\label{eq:Deltadef}
\end{equation}
Repeating this procedure for $X_{2}$ , $Y_{1}$ , and $Y_{2}$ and taking
$\Delta_{k}$ to be real we find that
\begin{eqnarray}
{\bf q} \cdot {\bf j}^{Q}_{2}({\bf q},\Omega) && =
- \sum_{k,\omega} (\Delta_{k+q} - \Delta_{k}) \nonumber\\
&& \times \left[\omega c_{k \uparrow}^{\dagger} c_{-(k+q) \downarrow}^{\dagger}
+ (\omega + \Omega) c_{-k \downarrow} c_{k+q \uparrow} \right]
\label{eq:qdotj2q}
\end{eqnarray}
In the limit as ${\bf q} \rightarrow 0$
\begin{equation}
\Delta_{k+q} - \Delta_{k} \approx {\bf q} \cdot
\frac{\partial \Delta_{k}}{\partial {\bf k}} \equiv {\bf q} \cdot
{\bf v}_{2}
\label{eq:v2def}
\end{equation}
Thus, casting (\ref{eq:qdotj2q}) in terms of the Nambu matrix formalism
we find that
\begin{eqnarray}
{\bf j}^{Q}_{2}(0,\Omega) = - \sum_{k,\omega} \left[
\left(\omega + \frac{\Omega}{2} \right) \right. && {\bf v}_{2}
\Psi_{k}^{\dagger} \tilde{\tau}_{1} \Psi_{k+q} \nonumber\\
&& + \left. \frac{\Omega}{2i} {\bf v}_{2}
\Psi_{k}^{\dagger} \tilde{\tau}_{2} \Psi_{k+q} \right] .
\label{eq:j2q}
\end{eqnarray}
In the limit of small $\Omega$, the second term can be neglected compared
to the first.  Thus, combining (\ref{eq:j2q}) with (\ref{eq:j1q}) we obtain
the following expression for the heat current in an anisotropic
superconductor:
\begin{eqnarray}
{\bf j}^{Q}(0,\Omega) = \sum_{k,\omega} \left(\omega+\frac{\Omega}{2}\right)
&& \left[ {\bf v}_{f} \Psi_{k}^{\dagger} \tilde{\tau}_{3} \Psi_{k+q}
\right. \nonumber\\
&& \left. - {\bf v}_{2} \Psi_{k}^{\dagger} \tilde{\tau}_{1} \Psi_{k+q}
\right] .
\label{eq:heatcurrent}
\end{eqnarray}
Note that for an s-wave superconductor, the gap is independent of ${\bf k}$
and ${\bf v}_{2} = 0$.  Thus, the second term in the heat current 
vanishes and (\ref{eq:heatcurrent}) reduces to the result derived by 
Ambegaokar {\it et al}. \cite{amb65,amb64}  However, for a
d-wave superconductor, the gap is anisotropic and ${\bf v}_{2} \neq 0$.
Hence, although the gap term may be small, neither term can be
formally neglected.

\subsection{Bare Bubble (Thermal)}
Unlike the electrical current,
the thermal current has two terms: a ``Fermi'' term proportional to 
${\bf v}_{f}$ and $\tilde{\tau}_{3}$ and a ``gap'' term proportional to
${\bf v}_{2}$ and $\tilde{\tau}_{1}$.  Therefore, when we evaluate
the current-current correlation function, we expect four bubbles rather
than just one: Fermi-Fermi, Fermi-gap, gap-Fermi, and gap-gap.  However,
since the Fermi velocity ${\bf v}_{f}$ and the gap velocity ${\bf v}_{2}$
are orthogonal at each of the gap nodes, the two cross terms cancel.
Hence, the thermal conductivity has two terms: a Fermi term with
velocity ${\bf v}_{f}$ and Nambu matrix $\tilde{\tau}_{3}$ on each vertex
and a gap term with velocity ${\bf v}_{2}$ and Nambu matrix
$\tilde{\tau}_{1}$ on each vertex.  For both terms the coupling parameter
is $(\omega + \Omega/2)$.

Neglecting vertex corrections, each term can be
obtained from the bare bubble generalized polarization function derived
in Appendix \ref{app:barebubble}.  Plugging the appropriate parameters
into the general result (\ref{eq:imretpolbare}) we find that
\begin{eqnarray}
&& \frac{\kappa (\Omega,T)}{T} = \frac{1}{\pi^{2}v_{f}v_{2}}
\int \frac{d^{2}p}{2\pi} \int_{-\infty}^{\infty} \!\!\!\!\!\! d\omega
\frac{n_{F}(\omega) - n_{F}(\omega+\Omega)}{\Omega} \nonumber \\
&& \;\;\;\; \times \left( \frac{\omega + \frac{\Omega}{2}}{T} \right)^{2}
\Biggl[ v_{f}^{2} \mbox{Tr} \left[ \tilde{G}''_{ret}({\bf p},\omega)
\tilde{\tau}_{3} \tilde{G}''_{ret}({\bf p},\omega+\Omega)
\tilde{\tau}_{3} \right] \Biggr. \nonumber \\
&& \;\;\;\;\;\;\;\;\;\;\;\;\;\;\;\;\;\;\;\;\;\;\;\;\,
\Biggl. +\, v_{2}^{2} \mbox{Tr}
\left[ \tilde{G}''_{ret}({\bf p},\omega)
\tilde{\tau}_{1} \tilde{G}''_{ret}({\bf p},\omega+\Omega)
\tilde{\tau}_{1} \right] \Biggr] \nonumber \\ &&
\label{eq:thermcondbare1}
\end{eqnarray}
In the universal limit ($\Omega\rightarrow0$, $T\rightarrow0$),
\begin{equation}
\frac{n_{F}(\omega) - n_{F}(\omega+\Omega)}{\Omega}
\rightarrow - \frac{\partial n_{F}}{\partial \omega}
\label{eq:diff2deriv}
\end{equation}
which for low $T$ is very sharply peaked at $\omega=0$.
Thus, evaluating the rest of the integrand for $\omega\rightarrow0$,
noting that $\Sigma_{ret}(0)=-i\Gamma_{0}$,
performing the frequency integral via
\begin{equation}
\int_{-\infty}^{\infty} \omega^{2}
\left( -\frac{\partial n_{F}}{\partial \omega} \right) d\omega
= \frac{\pi^{2}}{3} k_{B}^{2} T^{2}
\label{eq:intw2dndw}
\end{equation}
and integrating over momentum, we obtain the bare bubble thermal conductivity
in the universal limit:
\begin{equation}
\frac{\kappa_{0}}{T} = \left( \frac{\pi^{2}}{3} k_{B}^{2}
\right) \frac{1}{\pi^{2}} \frac{v_{f}^{2} + v_{2}^{2}}{v_{f} v_{2}} .
\label{eq:thermcondbare}
\end{equation}
Neglecting the $v_{2}^{2}$ term in the numerator,
this result and its finite temperature
corrections, were originally calculated by Graf {\it et al.\cite{gra96}}
The gap term was first obtained by Senthil {\it et al.\cite{sen98}}
via a physical argument
of Wiedemann-Franz correspondence with their expression for spin conductivity.
It arises here as a direct result of the additional gap term found in our
calculation of the thermal current for a d-wave superconductor.

\subsection{Vertex Corrections (Thermal)}
\label{ssec:thermvert}
The bare bubble result derived in the previous section can be
improved upon by including the contribution of the {\em ladder corrections}
to the bare vertex (see Fig.\ \ref{fig:vertcorrbub} of Appendix
\ref{app:vertexcorrections}).  A generalized polarization function including
such vertex corrections has been derived in Appendix
\ref{app:vertexcorrections}.  By plugging the appropriate parameters into
this general formula (\ref{eq:imretpolvert}) both terms of the thermal
conductivity (the Fermi term with parameters
${\bf v}_{f}$, $\tilde{\tau}_{3}$, and $\omega + \Omega/2$
and the gap term with parameters
${\bf v}_{2}$, $\tilde{\tau}_{1}$, and $\omega + \Omega/2$)
can be obtained.  Hence we find that
\begin{eqnarray}
\frac{\kappa(\Omega,T)}{T} && = \frac{1}{2 \pi^{2} v_{f} v_{2}}
\int_{-\infty}^{\infty} \!\!\!\!\!\! d\omega\
\frac{n_{F}(\omega) - n_{F}(\omega+\Omega)}{\Omega} \nonumber \\
&& \times \left( \frac{\omega + \frac{\Omega}{2}}{T} \right)^{2}
\Biggl[ v_{f}^{2} \mbox{Re} \left[ J^{(3)}_{2}(\omega,\Omega)
- J^{(3)}_{1}(\omega,\Omega) \right] \Biggr. \nonumber \\
&& \;\;\;\;\;\;\;\;\;\;\;\;\;\;\;\;\;\;\;\;\,
\Biggl. +\, v_{2}^{2} \mbox{Re}
\left[ J^{(1)}_{2}(\omega,\Omega) - J^{(1)}_{1}(\omega,\Omega)
\right] \Biggr] \nonumber \\ &&
\label{eq:thermcondvert1}
\end{eqnarray}
where $J^{\alpha}_{1}$ and $J^{\alpha}_{2}$ are defined in
appendix \ref{app:vertexcorrections}.
In the universal limit ($\Omega\rightarrow0$, $T\rightarrow0$),
the Fermi function factor is sharply peaked at $\omega=0$.
Thus, evaluating the J-functions for \mbox{$\Omega,\omega\rightarrow0$},
performing the frequency integral via (\ref{eq:intw2dndw}),
and noting that
\begin{eqnarray}
J^{(3)}_{2}(0,0) &=& -J^{(1)}_{1}(0,0) \nonumber \\
J^{(3)}_{1}(0,0) &=& -J^{(1)}_{2}(0,0)
\label{eq:J3J1}
\end{eqnarray}
we find that the universal limit thermal conductivity takes the form
\begin{equation}
\frac{\kappa_{0}}{T} = \left( \frac{\pi^{2}}{3} k_{B}^{2} \right)
\frac{1}{\pi^{2}} \frac{v_{f}^{2} + v_{2}^{2}}{v_{f} v_{2}} \beta^{T}_{vc}
\label{eq:thermcondvert2}
\end{equation}
\begin{equation}
\beta^{T}_{vc} = \frac{\frac{1}{2}}{1 - \gamma^{(0)}_{A2}}
+ \frac{\frac{1}{2}}{1 + \gamma^{(0)}_{A1} \left( 1 +
\frac{\gamma^{(0)}_{B1}}{\gamma^{(0)}_{A1}}
\frac{\gamma^{(0)}_{B1} \left( 2 \ln \frac{p_{0}}{\Gamma_{0}} - 2 \right)}
{1 - \gamma^{(0)}_{A1} \left( 2 \ln \frac{p_{0}}{\Gamma_{0}} - 2 \right)}
\right)}
\label{eq:betaVCthermal}
\end{equation}
where $\beta^{T}_{vc}$ is the thermal vertex correction factor and the
$\gamma$'s are defined in (\ref{eq:gammas0}).  As for the electrical case,
the thermal vertex correction must generally be evaluated numerically.
The results of such numerical calculations (presented in Appendix
\ref{app:numerical}) can be summarized as follows: (1) For all scattering
strengths (from Born to unitary) the thermal vertex correction is negligible
compared to the electrical vertex correction.  (2) In the small impurity
density limit, $\beta^{T}_{vc}-1$ vanishes approximately as
$[\ln(p_{0}/\Gamma_{0})]^{-1}$.
Thus, to zeroth order in the density of impurities,
vertex corrections do not contribute.  Hence,
\begin{equation}
\beta^{T}_{vc} \approx 1
\label{eq:betaVCthermalzero}
\end{equation}
and the universal limit thermal conductivity takes its bare bubble form
\begin{equation}
\frac{\kappa_{0}}{T} = \left( \frac{\pi^{2}}{3} k_{B}^{2} \right)
\frac{1}{\pi^{2}} \frac{v_{f}^{2} + v_{2}^{2}}{v_{f} v_{2}} .
\label{eq:thermcondvert}
\end{equation}
This is in stark contrast to the case of electrical conductivity
where we found a significant vertex correction
even to zeroth order in the impurity density.

\subsection{Fermi Liquid Corrections (Thermal)}
As discussed in Appendix \ref{app:fermiliquid}, there may be additional
corrections to the thermal conductivity due to underlying Fermi liquid
interactions between electrons.  In the $T\rightarrow0$ limit, the
dominant effect of such interactions is to renormalize the current
density operator.  From (\ref{eq:heatcurrent}) we see that the thermal
current (in the absence of Fermi liquid interactions)
has a rather complicated form including both a Fermi velocity term
and a gap velocity term.  However, for the purposes of this analysis,
it suffices to neglect the gap term (since it is known to be much smaller)
and note that in essence, the bare thermal current has the form
\begin{equation}
{\bf j}^{Q}_{0} = \sum_{k\alpha} \epsilon_{k} {\bf v}_{fk}
\,\delta n_{k\alpha}
\label{eq:barethermcurrent}
\end{equation}
where $\delta n_{k\alpha}$ is the deviation of the electron distribution
from equilibrium.  To account for the effects of the Fermi liquid interactions,
a general renormalization factor has been derived in Appendix \ref{app:fermiliquid}.
Plugging the appropriate parameters into (\ref{eq:dresscurrent}) we find that
the dressed thermal current is given by
\begin{eqnarray}
{\bf j}^{Q} = && {\bf j}^{Q}_{0} + \sum_{k'\alpha'} \delta n_{k'\alpha'}
\sum_{k} \epsilon_{k} {\bf v}_{fk}\, f^{s}_{kk'}
\Biggl[ \frac{\Delta_{k}^{2}}
{2 ( \epsilon_{k}^{2} + \Delta_{k}^{2} )^{3/2}} \nonumber \\
&& + \left( \frac{\epsilon_{k}^{2}}{E_{k}^{2}}
\frac{\Gamma_{0}/\pi}{E_{k}^{2} + \Gamma_{0}^{2}}
- \frac{\Delta_{k}^{2}}{\pi E_{k}^{3}} \arctan \left( \frac{\Gamma_{0}}{E_{k}}
\right) \right) \Biggr]
\label{eq:dressthermcurrent1}
\end{eqnarray}
where $f^{s}_{kk'}=f^{\uparrow\uparrow}_{kk'}+f^{\downarrow\uparrow}_{kk'}$.
Since the entire summand is odd in $\epsilon_{k}$, the correction
to the bare current integrates to zero.
Hence, at least in the zero temperature limit,
the thermal current is not renormalized by Fermi liquid effects.
\begin{equation}
{\bf j}^{Q} = {\bf j}^{Q}_{0}
\label{eq:dressthermcurrent}
\end{equation}
Basically, due to the symmetry of the electron dispersion about the Fermi
surface, corrections to the thermal current cancel.
Since the thermal current is unmodified, there are no Fermi liquid
corrections to the universal limit thermal conductivity.  Thus, including both
vertex corrections and Fermi liquid corrections, the thermal conductivity retains
the bare bubble form
\begin{equation}
\frac{\kappa_{0}}{T} = \left( \frac{\pi^{2}}{3} k_{B}^{2} \right)
\frac{1}{\pi^{2}} \frac{v_{f}^{2} + v_{2}^{2}}{v_{f} v_{2}} .
\label{eq:thermcondfermi}
\end{equation}

\section{Spin Conductivity}
\label{sec:spin}
For the spin conductivity case, the Kubo formula takes the form
\begin{equation}
\sigma^{s}(\Omega,T) = - \frac{\mbox{Im}\, \Pi^{s}_{ret}(\Omega)}{\Omega}
\label{eq:sconddef}
\end{equation}
where $\Pi^{s}_{ret}(\Omega)=\Pi^{s}(i\Omega \rightarrow \Omega
+ i\delta)$ and $\Pi^{s}(i\Omega)$ is the finite temperature
current-current correlation function (or polarization function).
Here the current that enters the correlation function is the spin
current derived below.

\subsection{Spin Current}
To find an expression for the spin current operator in an
anisotropic superconductor, we can write down the Hamiltonian and
spin density operators and use the spin continuity equation to
obtain the current.  For spin, the continuity equation is
\begin{equation}
\dot{\rho}^{s}({\bf x}) = - \nabla \cdot {\bf j}^{s}({\bf x})
\label{eq:spincont}
\end{equation}
where $\rho^{s}$ is the spin density and ${\bf j}^{s}$ is the spin
current density.  The spin density equation of motion takes the
standard form
\begin{equation}
\dot{\rho}^{s}({\bf x}) = -i \left[ \rho^{s}({\bf x}) , H \right] .
\label{eq:spineom}
\end{equation}
Thus, combining the two equations and using a Fourier representation for both
the spin density and the spin current we find that
\begin{equation}
{\bf q} \cdot {\bf j}^{s}_{q} = \left[ \rho^{s}_{q} , H \right] .
\label{eq:qdotj1}
\end{equation}
The Fourier transform of the spin density operator is given by
\begin{eqnarray}
\rho^{s}_{q} &=& \sum_{k,\omega,\alpha} S_{\alpha} c^{\dagger}_{k \alpha}
c_{k+q \alpha} \nonumber \\
&=& s \sum_{k,\omega} \left( c^{\dagger}_{k \uparrow} c_{k+q \uparrow}
- c^{\dagger}_{-k\downarrow} c_{-k+q \downarrow} \right)
\label{eq:rhoqdef}
\end{eqnarray}
where $S_{\alpha} = \pm \frac{1}{2}$ and $s \equiv \frac{1}{2}$.
In the mean field approximation, the Hamiltonian for a superconductor
is expressed as
\begin{eqnarray}
H = \sum_{k,\omega} && \left[ \epsilon_{k} \left( c^{\dagger}_{k\uparrow}
c_{k\uparrow} + c^{\dagger}_{-k\downarrow} c_{-k\downarrow} \right)
\right. \nonumber \\
&& \left. - \Delta_{k} \left( c^{\dagger}_{k\uparrow}
c^{\dagger}_{-k\downarrow} + c_{-k\downarrow} c_{k\uparrow} \right) \right] .
\label{eq:HamMF}
\end{eqnarray}
Thus, evaluating the commutator of (\ref{eq:rhoqdef}) and (\ref{eq:HamMF})
using fermionic anticommutation relations we obtain
\begin{eqnarray}
{\bf q} \cdot {\bf j}^{s}_{q} && = s \sum_{k,\omega} \left[
(\epsilon_{k+q} - \epsilon_{k}) \left( c^{\dagger}_{k\uparrow} c_{k+q\uparrow}
- c_{-k\downarrow} c^{\dagger}_{-(k+q)\downarrow} \right)
\right. \nonumber \\
&& \left. - (\Delta_{k+q} - \Delta_{k}) \left( c^{\dagger}_{k\uparrow}
c^{\dagger}_{-(k+q)\downarrow} + c_{-k\downarrow} c_{k+q\uparrow} \right)
\right] .
\label{eq:qdotj2}
\end{eqnarray}
In the limit as $q \rightarrow 0$
\begin{equation}
\epsilon_{k+q} - \epsilon_{k} \approx {\bf q} \cdot
\frac{\partial \epsilon_{k}}{\partial {\bf k}} \equiv {\bf q} \cdot
{\bf v}_{f}
\label{eq:epslimit}
\end{equation}
\begin{equation}
\Delta_{k+q} - \Delta_{k} \approx {\bf q} \cdot
\frac{\partial \Delta_{k}}{\partial {\bf k}} \equiv {\bf q} \cdot
{\bf v}_{2}
\label{eq:deltalimit}
\end{equation}
Hence, expressing the creation and annihilation operators in terms of
2x2 Nambu matrix notation we find that
\begin{equation}
{\bf j}^{s}(0,\Omega) = s \sum_{k,\omega} \left[ {\bf v}_{f}
\Psi^{\dagger}_{k} \tilde{\tau}_{3} \Psi_{k+q}
- {\bf v}_{2} \Psi^{\dagger}_{k} \tilde{\tau}_{1} \Psi_{k+q} \right] .
\label{eq:spincurrent}
\end{equation}
Note that the spin current takes precisely the same form as the thermal
current (\ref{eq:heatcurrent}) with an appropriate change of coupling parameter.

\subsection{Bare Bubble (Spin)}
As in the thermal conductivity
case, the spin current has both a ``Fermi'' term and a ``gap'' term.
Hence, evaluating the current-current correlation function and noting
(as before) that the cross-terms cancel, we find that the spin conductivity
consists of two bubbles: a Fermi term with ${\bf v}_{f}$ and $\tilde{\tau}_{3}$
on each vertex and a gap term with ${\bf v}_{2}$ and $\tilde{\tau}_{1}$
on each vertex.  These are precisely the bubbles that we evaluated for 
the thermal case except that here the coupling constant is the spin ($s=1/2$)
rather than the frequency.

Neglecting vertex corrections, each of the two bubbles can be evaluated by
plugging the appropriate set of parameters into the bare bubble
generalized polarization function (\ref{eq:imretpolbare}) derived in
Appendix \ref{app:barebubble}.  Doing so we find that
\begin{eqnarray}
\sigma^{s}(\Omega,T) && = \frac{s^{2}}{\pi^{2}v_{f}v_{2}}
\int \frac{d^{2}p}{2\pi} \int_{-\infty}^{\infty} \!\!\!\!\!\! d\omega
\frac{n_{F}(\omega) - n_{F}(\omega+\Omega)}{\Omega} \nonumber \\
&& \times \Biggl[ v_{f}^{2} \mbox{Tr} \left[ \tilde{G}''_{ret}({\bf p},\omega)
\tilde{\tau}_{3} \tilde{G}''_{ret}({\bf p},\omega+\Omega)
\tilde{\tau}_{3} \right] \Biggr. \nonumber \\
&& \Biggl. +\, v_{2}^{2} \mbox{Tr}
\left[ \tilde{G}''_{ret}({\bf p},\omega)
\tilde{\tau}_{1} \tilde{G}''_{ret}({\bf p},\omega+\Omega)
\tilde{\tau}_{1} \right] \Biggr] .
\label{eq:spincondbare1}
\end{eqnarray}
In the universal limit ($\Omega\rightarrow0$, $T\rightarrow0$),
\begin{equation}
\frac{n_{F}(\omega) - n_{F}(\omega+\Omega)}{\Omega}
\rightarrow - \frac{\partial n_{F}}{\partial \omega}
\rightarrow \delta(\omega) .
\label{eq:spindiff2deriv2delta}
\end{equation}
Thus, evaluating the rest of the integrand in the \mbox{$\omega\rightarrow0$}
limit, noting that $\Sigma_{ret}(0)=-i\Gamma_{0}$,
and integrating over momentum, we obtain the universal limit
bare bubble spin conductivity:
\begin{equation}
\sigma^{s}_{0} = \frac{s^{2}}{\pi^{2}}
\frac{v_{f}^{2} + v_{2}^{2}}{v_{f} v_{2}} .
\label{eq:spincondbare}
\end{equation}
This agrees (aside from a disputed factor of 2) with the
result obtained by Senthil {\it et al.} \cite{sen98}

\subsection{Vertex Corrections (Spin)}
\label{ssec:spinvert}
As in the electrical and thermal cases discussed previously, the
bare bubble result derived above can be improved upon by
including the contribution of the ladder corrections to the
bare vertex.  By plugging the appropriate parameters into the
generalized polarization function (including vertex corrections)
derived in Appendix \ref{app:vertexcorrections}, both the Fermi
term and the gap term of the spin conductivity can be obtained.
Doing so we find that
\begin{eqnarray}
\sigma^{s}(\Omega,T) = && \frac{s^{2}}{2 \pi^{2} v_{f} v_{2}}
\int_{-\infty}^{\infty} \!\!\!\!\!\! d\omega\
\frac{n_{F}(\omega) - n_{F}(\omega+\Omega)}{\Omega} \nonumber \\
&& \times \Biggl[ v_{f}^{2} \mbox{Re} \left[ J^{(3)}_{2}(\omega,\Omega)
- J^{(3)}_{1}(\omega,\Omega) \right] \Biggr. \nonumber \\
&& \,\ \Biggl. +\, v_{2}^{2} \mbox{Re}
\left[ J^{(1)}_{2}(\omega,\Omega) - J^{(1)}_{1}(\omega,\Omega)
\right] \Biggr]
\label{eq:spincondvert1}
\end{eqnarray}
where $J^{\alpha}_{1}$ and $J^{\alpha}_{2}$ are defined in
Appendix \ref{app:vertexcorrections}.
In the universal limit ($\Omega\rightarrow0$, $T\rightarrow0$),
we can make use of (\ref{eq:spindiff2deriv2delta}) to evaluate the
frequency integral and find that
\begin{equation}
\sigma^{s}_{0} = \frac{s^{2}}{\pi^{2}}
\frac{v_{f}^2 + v_{2}^{2}}{v_{f} v_{2}} \beta^{s}_{vc}
\label{eq:spincondvert2}
\end{equation}
where the spin vertex correction factor, $\beta^{s}_{vc}$, is identical
to the thermal vertex correction factor, $\beta^{T}_{vc}$, defined
in (\ref{eq:betaVCthermal}).  Thus, mirroring the analysis described in
Sec.\ \ref{ssec:thermvert}, we note that for small impurity density,
$\beta^{s}_{vc} \approx 1$ and the universal limit spin conductivity
takes its bare bubble form
\begin{equation}
\sigma^{s}_{0} = \frac{s^{2}}{\pi^{2}}
\frac{v_{f}^2 + v_{2}^{2}}{v_{f} v_{2}} .
\label{eq:spincondvert}
\end{equation}
As in the thermal case and in contrast to
the electrical case, vertex corrections do not contribute
(to zeroth order in the impurity density).

\subsection{Fermi Liquid Corrections (Spin)}
An additional correction to the spin conductivity is needed to account
for the effects of underlying Fermi liquid interactions between electrons.
In the $T\rightarrow0$ limit, the dominant effect of Fermi liquid
interactions is the renormalization of the current density operator
(see Appendix \ref{app:fermiliquid}).
Neglecting the gap term in (\ref{eq:spincurrent}) (since it is small
and rather difficult to deal with), the spin current has the basic form
\begin{equation}
{\bf j}^{s}_{0} = \sum_{k\alpha} S_{\alpha} {\bf v}_{fk}
\,\delta n_{k\alpha}
\label{eq:barespincurrent}
\end{equation}
where $S_{\alpha}=\pm1/2$.
The Fermi liquid renormalization of this current can then be obtained 
by plugging the appropriate parameters into the general result
(\ref{eq:dresscurrent}) derived in Appendix \ref{app:fermiliquid}.
Noting that the math is completely analogous to that of the
electrical calculation in Sec.\ \ref{ssec:elecfermi}, we can easily
adapt the electrical result to the present case.  Replacing
$-e$ with $s$ $(=1/2)$ and changing spin-symmetric designations
to spin-antisymmetric ones, we find that the Fermi liquid renormalization
of the spin current takes the form
\begin{equation}
{\bf j}^{s} = {\bf j}^{s}_{0} \alpha_{fl}^{a}
\label{eq:dressspincurrent}
\end{equation}
where $\alpha_{fl}^{a}$ is the spin-antisymmetric current renormalization
factor which, for a general Fermi surface, is some complicated function
of the spin-antisymmetric Landau parameters, $F^{a}_{\ell}$.  For
the simplified case of a circular Fermi surface,
\begin{equation}
\alpha_{fl}^{a} \approx 1 + \frac{F^{a}_{1}}{2} .
\label{eq:alphaspincircular}
\end{equation}

As in the electrical case, the current renormalization is dominated
by a Fermi surface term resulting from the interaction-induced modification
of the equilibrium distribution of the condensate.  At first glance, this
result is a bit surprising since the Zeeman field which generates the
(normal) spin current cannot induce a supercurrent.  However,
it must be understood that the renormalization of the
normal current has nothing to do with the presence of a supercurrent.  Rather,
due to the existence of the normal current, the {\em equilibrium} distribution
of the condensate is modified (via Fermi liquid effects).  It is this modification
of the equilibrium condensate, not the presence of an excited
condensate (supercurrent), that gives rise to the renormalization of the
spin current.

Since the spin conductivity is proportional to the current-current correlation
function, it obtains two factors of the spin current renormalization.
Thus, including both vertex corrections and Fermi liquid corrections, the
universal limit spin conductivity takes the form
\begin{equation}
\sigma^{s}_{0} = \frac{s^{2}}{\pi^{2}}
\frac{v_{f}^2 + v_{2}^{2}}{v_{f} v_{2}}
\left. \alpha_{fl}^{a} \right.^{2} .
\label{eq:spincondfermi}
\end{equation}

\section{Conclusions}
\label{sec:conclusions}
In the presence of impurities, the gap symmetry of a d-wave superconductor
yields the generation of impurity-induced quasiparticles at the gap nodes.
The transport properties of the resulting system are quite unique since such
quasiparticles are both generated and scattered by impurities.  In the
$\Omega\rightarrow0$, $T\rightarrow0$ limit, bare bubble calculations
indicate that transport coefficients are ``universal'', independent of
the impurity density or scattering rate.
However, once the contributions of vertex corrections and Fermi liquid
corrections are included, we find that (putting in the $\hbar$'s) the
electrical, thermal, and spin conductivities in this universal limit take
the form
\begin{mathletters}
\label{eq:conductivities}
\begin{eqnarray}
\sigma_{0} &=& \frac{e^{2}}{\hbar \pi^{2}} \frac{v_{f}}{v_{2}} \beta_{vc}
\left. \alpha_{fl}^{s} \right.^{2}
\\
\frac{\kappa_{0}}{T} &=& \frac{\left( \frac{\pi^{2}}{3} k_{B}^{2} \right)}
{\hbar \pi^{2}} \left( \frac{v_{f}}{v_{2}} + \frac{v_{2}}{v_{f}} \right)
\\
\sigma^{s}_{0} &=& \frac{s^{2}}{\hbar \pi^{2}}
\left( \frac{v_{f}}{v_{2}} + \frac{v_{2}}{v_{f}} \right)
\left. \alpha_{fl}^{a} \right.^{2}
\end{eqnarray}
\end{mathletters}
where $\beta_{vc}$ is a scattering dependent vertex correction
(\ref{eq:betaVC}) and $\alpha_{fl}^{s}$ and $\alpha_{fl}^{a}$ are
spin-symmetric and spin-antisymmetric Fermi liquid factors
(\ref{eq:dresseleccurrent},\ref{eq:dressspincurrent}).  Note that
these are the 2D conductivities of a single CuO$_{2}$ plane.  To obtain
3D conductivities, they must be multiplied by the number of CuO$_{2}$
planes per unit length stacked along the c-axis.

The ``law'' of Wiedemann and Franz suggests that the
transport coefficients should be related such that
\begin{equation}
\frac{\kappa}{\sigma T} = \frac{\pi^{2}}{3} \frac{k_{B}^{2}}{e^{2}}
\;\;\;\;\;\;\;\;
\frac{\kappa}{\sigma^{s} T} = \frac{\pi^{2}}{3} \frac{k_{B}^{2}}{s^{2}}
\;\;\;\;\;\;\;\;
\frac{\sigma^{s}}{\sigma} = \frac{s^{2}}{e^{2}} .
\label{eq:WiedemannFranz}
\end{equation}
However, examination of the expressions above yields three sources of
Wiedemann-Franz violation: current operator definition corrections, vertex
corrections, and Fermi liquid corrections.
First of all, since the electrical current has only a Fermi term
while the thermal and spin currents include both a Fermi term and a gap term,
$\sigma_{0}$ is proportional to the ratio $v_{f}/v_{2}$ while $\kappa_{0}$
and $\sigma^{s}_{0}$ involve an extra $v_{2}/v_{f}$ term.
These extra terms arise when the thermal and spin current operators
are corrected to account for the anisotropy of the order parameter.
However, since $v_{f}/v_{2} \sim 14$ for YBCO
\cite{chi99}, this type of violation is of more qualitative than
quantitative importance.
Secondly, unless impurity scattering is completely isotropic in k-space,
the electrical conductivity contains a scattering dependent vertex
correction, $\beta_{vc}$, which cannot be neglected even to zeroth order
in impurity density.  However, analogous corrections to the
thermal and spin conductivities
vanish in the small impurity density limit.  Thus we expect a scattering
dependent enhancement of $\sigma_{0}$ that is absent in $\kappa_{0}$ and
$\sigma^{s}_{0}$.
Finally, due to underlying Fermi liquid interactions, the electrical and
spin conductivities gain spin-symmetric and spin-antisymmetric
correction factors respectively.
Corresponding corrections to the thermal current
cancel due to particle-hole symmetry.  Hence, while Fermi-liquid
interactions modify $\sigma_{0}$ and $\sigma^{s}_{0}$, the value of
$\kappa_{0}$ is unaffected.

\begin{figure}[tb]
\centerline{\psfig{file=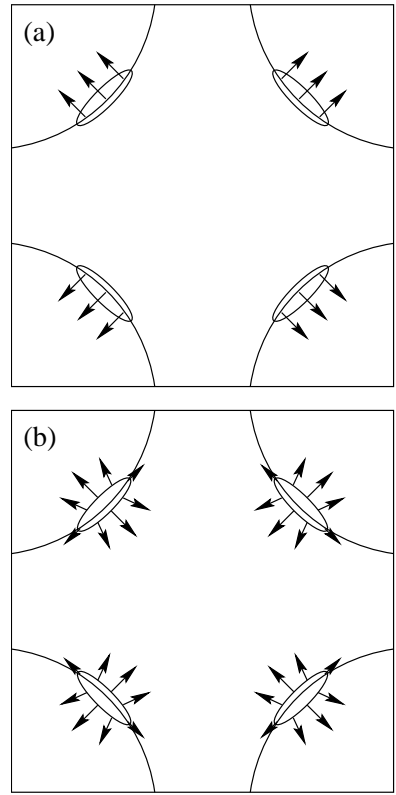}}
\vspace{0.5 cm}
\caption{Schematic depictions of the (a) electrical current and
(b) thermal/spin current in the vicinity of the four gap nodes.  Electrical
current is proportional to the Fermi velocity,
${\bf v}_{f}=\partial \epsilon_{k} / \partial {\bf k}$, whereas
thermal/spin current is proportional to the group velocity,
${\bf v}_{G}=\partial E_{k} / \partial {\bf k}$.  The ellipses drawn at each
node denote (at a very exaggerated scale) the regions of k-space within which
impurity-induced quasiparticles are generated in the universal limit. (Note
that in the small impurity density regime with which we are concerned
($\Gamma_{0}\ll\Delta_{0}$), these nodal regions are point-like
on the scale of the Brillouin zone.)}
\label{fig:velocities}
\end{figure}

The physical origin of the first two corrections lies with the
velocity dependence of the current operators.  Although somewhat obscured
in the Nambu formalism, when our current operators
(\ref{eq:elecjqnambu},\ref{eq:heatcurrent},\ref{eq:spincurrent})
are rewritten in the quasiparticle basis, it is clear that the
electrical current is proportional to the Fermi velocity,
${\bf v}_{f}=\partial \epsilon_{k} / \partial {\bf k}$, while the thermal
and spin currents are proportional to the group velocity,
${\bf v}_{G}=\partial E_{k} / \partial {\bf k}$.
This difference arises because quasiparticles carry definite energy and
spin but do not carry definite charge.
Since energy and spin are well defined in the quasiparticle basis,
thermal and spin currents are proportional to the group velocity,
the derivative of the quasiparticle dispersion.
By contrast, the electron and hole parts of each quasiparticle have
opposite charge and opposite velocity.  Therefore each part carries
the same electrical current, proportional to the normal state Fermi velocity.
This point was emphasized in Ref.\ \onlinecite{lee97}.
For a d-wave superconductor where both $\epsilon_{k}$ and $\Delta_{k}$
are momentum-dependent, the group velocity will have both a ${\bf v}_{f}$
component and a ${\bf v}_{2}$ component while the Fermi velocity can only
have a ${\bf v}_{f}$ component (see Fig.\ \ref{fig:velocities}).
This is the source of the extra gap terms in the thermal and spin conductivities.
(Similar conclusions were drawn by Moreno and Coleman. \cite{mor96})

The role of vertex corrections can
be understood by considering the graphical depictions of the Fermi velocity
and group velocity presented in Fig.\ \ref{fig:velocities}.
Throughout the area of a node, the magnitude and direction of the
Fermi velocity is approximately constant.  Thus, the electrical
current can relax much more effectively via scattering from node to node than
it can via scattering within a single node.  It is therefore
necessary to distinguish, mathematically, between the effects of intra-node
scattering and inter-node scattering.  This is accomplished through the
inclusion of vertex corrections.  In contrast, the group velocity varies
significantly over the area of a node.  Therefore, the thermal and spin currents
can relax through either intra-node scattering or scattering between nodes.
As a result, the different types of scattering play nearly the same role
and need not be distinguished.  Hence, vertex corrections do not contribute
to the thermal and spin conductivity.

Fermi liquid corrections result from the redistribution of equilibrium electrons
in response to the presence of interactions between excited electrons.
In essence, this redistribution gives rise to a drag current that can renormalize
the quasiparticle current and therefore the conductivity.
The character of the renormalization depends on the nature of the coupling parameter
for a particular current.  Since the spin current gets opposite contributions from
the two species of spin, the spin conductivity gets a spin-antisymmetric
renormalization.  However, charge is spin-independent so the electrical conductivity
gets a spin-symmetric renormalization.
Furthermore, since energy changes sign across the Fermi surface, particle-hole
symmetry dictates that the effects of Fermi liquid interactions on
the thermal current must cancel.  Thus, thermal conductivity is not renormalized.

The velocity ratio, $v_{f}/v_{2}$, is a fundamental material parameter
which measures the anisotropy of the quasiparticle excitation spectrum.
Therefore, an important objective in measuring
quantities such as the normal fluid density
and the universal limit transport coefficients,
which all depend on $v_{f}/v_{2}$, is to obtain the value of this ratio.
However, due to vertex corrections and Fermi liquid corrections, the
electrical conductivity, spin conductivity, and normal fluid density
depend on parameters (such as interaction energy and/or
scattering potential) with values that are not well known.  Only
the thermal conductivity involves neither vertex corrections nor
Fermi liquid corrections.  Thus, $\kappa_{0}$ is the only
truly ``universal'' coefficient and is the quantity from which the
value of $v_{f}/v_{2}$ can be most directly obtained.
On the other hand, the linear $T$ coefficient of the superfluid density
(\ref{eq:rhofermi}) is proportional to
$\left. \alpha_{fl}^{s} \right.^{2} v_{f}/v_{2}$.
Hence, these two measurements can be combined to determine the Fermi
liquid factor, $\alpha_{fl}^{s}$.

In fact, while this paper was in preparation, Chiao {\it et al.\/}
\cite{chi99A} applied these conclusions to the results of a series of
recent experiments performed on optimally-doped
Bi$_{2}$Sr$_{2}$CaCu$_{2}$O$_{8}$ (BSCCO)\@.  By analyzing the residual
linear term in their very low temperature thermal conductivity measurements
in terms of (\ref{eq:conductivities}b), they obtained a value for the
velocity ratio, $v_{f}/v_{2}=20$.  This is the same value obtained from
the ARPES measurements of Mesot {\it et al.} \cite{mes99}  Going further,
by combining this result with the linear $T$ coefficient of the
superfluid density measured by Waldram and co-workers \cite{wal96} and
making use of (\ref{eq:rhofermi}), they extracted a value for the
Fermi-liquid correction, $\left. \alpha_{fl}^{s} \right.^{2}=0.41$.
These results provide an experimental verification of our analysis.

\acknowledgments
The authors gratefully acknowledge discussions with A. J. Berlinsky, C. Kallin,
D. Bonn, and L. Taillefer and helpful comments from M. R. Norman and J. A. Sauls.
A.~C.~D. thanks A. Abanov and M. Oktel for valuable
help with calculations.  This material is based upon work supported under a
National Science Foundation Graduate Fellowship as well as NSF grant
DMR-9813764.

\appendix

\section{Bare Bubble Calculation}
\label{app:barebubble}
\begin{figure}[b]
\centerline{\psfig{file=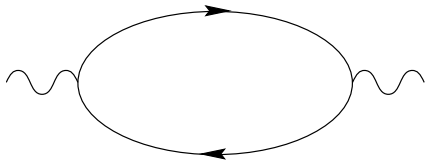}}
\vspace{0.5 cm}
\caption{Bare polarization bubble.}
\label{fig:barebub}
\end{figure}

In the absence of vertex corrections, our calculations of electrical,
thermal, and spin conductivity all require the evaluation of {\em bare bubble}
diagrams depicting various types of polarization functions.
The details of these different calculations are all quite similar.  They
differ only in the coupling parameter, velocity, and Pauli matrix
contributed by each bare vertex.  Rather than repeating the same basic
calculation several times, it is convenient to calculate a generalized
polarization function here which can be referred to for
each of the specific cases of interest.  This generalized
function $\tensor{\Pi}^{g\ell\alpha}$,
will depend on a coupling parameter g, a velocity ${\bf v}_{\ell}$,
and a Nambu space Pauli matrix $\tilde{\tau}_{\alpha}$ where
\begin{eqnarray*}
g &=& \{ e, s(=1/2), \omega + \Omega/2 \} \\
{\bf v}_{\ell} &=& \{ {\bf v}_{f}, {\bf v}_{2} \} \\
\tilde{\tau}_{\alpha} &=& \{ \tilde{\tau}_{0}(=\tilde{\openone}),
\tilde{\tau}_{1}, \tilde{\tau}_{2}, \tilde{\tau}_{3} \} .
\end{eqnarray*}
Evaluating the diagram in Fig.\ \ref{fig:barebub} we find that
\begin{equation}
\tensor{\Pi}^{g\ell\alpha}(i\Omega) = \frac{1}{\beta}\sum_{k, i\omega}
g^{2} {\bf v}_{\ell} {\bf v}_{\ell} \mbox{Tr} \Bigl[
\tilde{\cal{G}}({\bf k},i\omega) \tilde{\tau}_{\alpha}
\tilde{\cal{G}}({\bf k},i\omega + i\Omega) \tilde{\tau}_{\alpha}
\Bigr]
\label{eq:polbare1}
\end{equation}
where $\tilde{\cal{G}}({\bf k},i\omega)$ is the $2 \times 2$
Nambu matrix form of the Matsubara Green's function.
For a d-wave superconductor at temperatures much less than the gap 
maximum, quasiparticles are generated primarily at the four gap nodes.
\cite{lee93,lee97}
Thus, linearizing the quasiparticle spectrum about the nodes and defining
a coordinate system $(k_{1},k_{2})$ at each node with $\hat{\bf k}_{1}$
($\hat{\bf k}_{2}$) perpendicular (parallel) to the
Fermi surface, we can replace our momentum sum by
an integral over the k-space area surrounding each node.  If we further
define a scaled momentum $(p_{1},p_{2})$ we can let
\begin{equation}
\sum_{k} \rightarrow \sum_{j=1}^{4}\int\frac{d^{2}k}{(2\pi)^{2}}
\rightarrow \sum_{j=1}^{4}\int\frac{d^{2}p}{(2\pi)^{2}v_{f}v_{2}}
\label{eq:sum2int}
\end{equation}
where $p_{1} \equiv v_{f}k_{1} = \varepsilon_{{\bf k}}$ and
$p_{2} \equiv v_{2}k_{2} = \Delta_{{\bf k}}$.  Since
${\bf v}_{f} = v_{f}\hat{\bf k}_{1}$ and ${\bf v}_{2} = v_{2}\hat{\bf k}_{2}$
at each of the four nodes, the sum over nodes yields
\begin{equation}
\sum_{j=1}^{4} {\bf v}_{\ell}^{(j)} {\bf v}_{\ell}^{(j)}
= 2v_{\ell}^{2} \tensor{\openone} .
\label{eq:vvsum}
\end{equation}
This, in turn, allows the definition of a scalar polarization function via
\begin{equation}
\tensor{\Pi}^{g\ell\alpha}(i\Omega) \equiv \Pi^{g\ell\alpha}(i\Omega)
\tensor{\openone} .
\label{eq:tensor2scalar}
\end{equation}
Defining a spectral representation for $\tilde{\cal{G}}$,
we can write
\begin{equation}
\tilde{\cal{G}}({\bf p},i\omega) = \int_{-\infty}^{\infty}
\frac{\tilde{A}({\bf p},\omega_{1})}{i\omega - \omega_{1}}d\omega_{1}
\label{eq:specrep}
\end{equation}
where
\begin{equation}
\tilde{A}({\bf p},\omega) = -\frac{1}{\pi}
\tilde{G}''_{ret}({\bf p},\omega)
\label{eq:specfunc}
\end{equation}
and $\tilde{G}''_{ret}$ is the imaginary part of the retarded
Green's function.  Plugging back into (\ref{eq:polbare1}) we obtain
\begin{eqnarray}
\Pi^{g\ell\alpha}(i\Omega) = && \frac{2v_{\ell}^{2}}{v_{f}v_{2}}
\int \frac{d^{2}p}{(2\pi)^{2}} \int d\omega_{1} \int d\omega_{2}
\nonumber\\
&& \times \mbox{Tr} \Bigl[ \tilde{A}({\bf p},\omega_{1})
\tilde{\tau}_{\alpha} \tilde{A}({\bf p},\omega_{2})
\tilde{\tau}_{\alpha} \Bigr] S
\label{eq:polbare2}
\end{eqnarray}
where
\begin{equation}
S = \frac{1}{\beta} \sum_{i\omega} g^{2} \frac{1}{i\omega - \omega_{1}}
\frac{1}{i\omega + i\Omega - \omega_{2}} .
\label{eq:Sbare}
\end{equation}
Evaluating the Matsubara sum in the standard way \cite{mah90}
we pick up a contribution from each of the poles of the summand.
Since the intermediate results differ depending on the frequency-dependence
of the coupling parameter, it is best to handle the frequency-independent
coupling and frequency-dependent coupling cases separately.

For $g = \{e,s\}$ (frequency-independent coupling), the sum is 
straightforward.  Adding the contribution of the two poles and then
continuing $i\Omega \rightarrow \Omega + i\delta$
we obtain the retarded function
\begin{equation}
S_{ret} = S(i\Omega \rightarrow \Omega + i\delta)
= g^{2} \frac{n_{F}(\omega_{1}) - n_{F}(\omega_{2})}%
{\omega_{1} - \omega_{2} + \Omega + i\delta}
\label{eq:Sfreqind}
\end{equation}
where
\begin{equation}
n_{F}(\omega) = \frac{1}{e^{\beta\omega} + 1}
\end{equation}
is the Fermi function.

For $g = \omega + \Omega/2$ (frequency-dependent coupling),
we proceed in the same way but there
are a few technicalities that must be clarified.  First of all, it
should be understood that within the Matsubara sum we really mean 
$g \rightarrow i\omega + i\Omega/2$.  Only after the sum has been
evaluated and all frequencies have been continued to the real axis
should the stated form of the coupling parameter be taken literally.
Secondly, note that with this frequency dependent g,
the summand has two extra powers of frequency.  As a result, the sum
appears to be divergent.  However, as discussed by Ambegaokar and 
Griffin \cite{amb65}, this apparent divergence results from an improper
treatment of time-derivatives within the time-ordered correlation 
function and should be ignored.  Doing so, we proceed just as before.
Adding the contribution of the two summand poles and continuing the
external frequency to the real axis we find that
\begin{equation}
S_{ret} = \frac{(\omega_{1} + \frac{\Omega}{2})^{2} n_{F}(\omega_{1})
- (\omega_{2} - \frac{\Omega}{2})^{2} n_{F}(\omega_{2})}
{\omega_{1} - \omega_{2} + \Omega + i\delta} .
\label{eq:Sfreqdep}
\end{equation}

Plugging (\ref{eq:Sfreqind}) and (\ref{eq:Sfreqdep}) back into
(\ref{eq:polbare2}), writing the spectral function in terms of the
retarded Green's function via (\ref{eq:specfunc}), and  using the identity
\begin{equation}
\frac{1}{x + i\delta} = \mbox{P}\frac{1}{x} - i\pi\delta(x)
\end{equation}
to take the imaginary part, we find that
\begin{eqnarray}
\mbox{Im}\, \Pi_{ret}^{g\ell\alpha}(\Omega) &&=  \frac{1}{\pi^{2}}
\frac{v_{\ell}^{2}}{v_{f}v_{2}}
\int \frac{d^{2}p}{2\pi} \int_{-\infty}^{\infty} d\omega \nonumber\\
&& \times g^{2} \left( n_{F}(\omega+\Omega) - n_{F}(\omega) \right) \nonumber\\
&& \times \mbox{Tr} \Bigl[ \tilde{G}''_{ret}({\bf p},\omega)
\tilde{\tau}_{\alpha} \tilde{G}''_{ret}({\bf p},\omega+\Omega)
\tilde{\tau}_{\alpha} \Bigr]
\label{eq:imretpolbare}
\end{eqnarray}
for all three coupling parameters $g = \{e,s,\omega+\Omega/2\}$.
Neglecting vertex corrections, this is the imaginary part
of the generalized retarded polarization function.
The real part can be obtained via Kramers-Kronig analysis.
By specifying different input parameters, (\ref{eq:imretpolbare}) can
be used to obtain the electrical, thermal, and spin conductivity.

\section{Vertex Corrections}
\label{app:vertexcorrections}
\begin{figure}[tb]
\centerline{\psfig{file=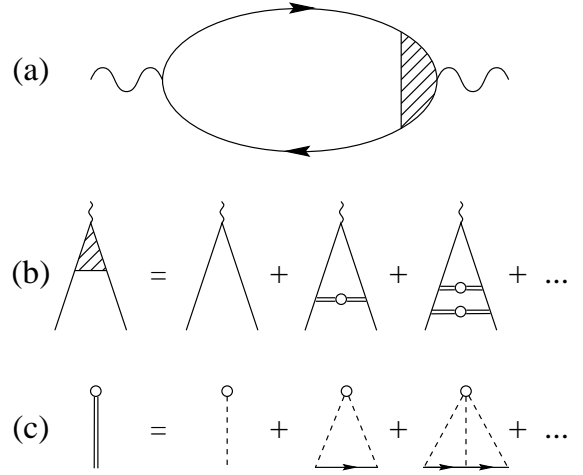}}
\vspace{0.5 cm}
\caption{(a) Polarization bubble with dressed vertex.  (b) Ladder series.
(c) T-matrix series.}
\label{fig:vertcorrbub}
\end{figure}

Unless the scattering potential is completely isotropic, the bare bubble
results of Appendix \ref{app:barebubble} can be improved upon by including
the contributions of vertex corrections.  In this section the
{\em ladder corrections} depicted in Fig.\ \ref{fig:vertcorrbub} will be
included.  Once again, our object is to obtain an expression for a generalized
polarization function, $\tensor{\Pi}^{g\ell\alpha}$,
in which each of the vertices contribute a
coupling parameter g, a velocity ${\bf v}_{\ell}$, and a Nambu space Pauli
matrix $\tilde{\tau}_{\alpha}$.
Evaluating the diagram in Fig.\ \ref{fig:vertcorrbub}(a) and noting that
${\bf v}_{\ell} \equiv v_{\ell} \hat{\bf k}_{\ell}$
we find that the generalized polarization function takes the form
\begin{eqnarray}
\tensor{\Pi}^{g\ell\alpha} && (i\Omega) = \frac{1}{\beta}\sum_{i\omega}\sum_{k}
g^{2} v_{\ell}^{2} \hat{\bf k}_{\ell} \nonumber\\
&& \times \mbox{Tr} \Bigl[
\tilde{\cal{G}}({\bf k},i\omega) \tilde{\tau}_{\alpha}
\tilde{\cal{G}}({\bf k},i\omega + i\Omega) \tilde{\tau}_{\alpha}
\tilde{\bf \Gamma}^{\ell\alpha}({\bf k},i\omega,i\Omega) \Bigr]
\label{eq:polvert1}
\end{eqnarray}
which is equivalent to the bare bubble result (\ref{eq:polbare1}) with the
unit vector $\hat{\bf k}_{\ell}$ from the second bare vertex replaced
by a more general vertex function $\tilde{\bf \Gamma}^{\ell\alpha}$.
Evaluating the diagram series in Fig.\ \ref{fig:vertcorrbub}(b) we
obtain an equation which can be solved for the vertex function:
\begin{eqnarray}
\tilde{\tau}_{\alpha} \tilde{\bf \Gamma}^{\ell\alpha}({\bf k})
= \hat{\bf k}_{\ell} \tilde{\tau}_{\alpha} + && n_{i} \sum_{k''}
\tilde{T}_{kk''}(i\omega+i\Omega)
\tilde{\cal{G}}({\bf k}'',i\omega + i\Omega) \nonumber\\
&& \times \tilde{\tau}_{\alpha}
\tilde{\bf \Gamma}^{\ell\alpha}({\bf k}'')
\tilde{\cal{G}}({\bf k}'',i\omega)
\tilde{T}_{k''k}(i\omega)
\label{eq:eqforGamma}
\end{eqnarray}
where $n_{i}$ is the impurity density and
$\tilde{T}_{kk'}(i\omega)$ is the impurity scattering T-matrix defined
by the diagram series in Fig.\ \ref{fig:vertcorrbub}(c).
Multiplying from the left by $\tilde{\tau}_{\alpha}$ we can define
\begin{equation}
\tilde{\Gamma}^{\ell\alpha}({\bf k},i\omega,i\Omega)
= \hat{\bf k}_{\ell} \left[ \tilde{\openone}
+ \tilde{\Lambda}^{\alpha}(i\omega,i\Omega) \right]
\label{eq:GammafromLambda}
\end{equation}
where
\begin{eqnarray}
\hat{\bf k}_{\ell} \tilde{\Lambda}^{\alpha} = n_{i} \sum_{k''}
&& \tilde{\tau}_{\alpha} \tilde{T}_{kk''}(i\omega+i\Omega)
\tilde{\cal{G}}({\bf k}'',i\omega + i\Omega)
\tilde{\tau}_{\alpha} \nonumber\\
&& \times \tilde{\bf \Gamma}^{\ell\alpha}({\bf k}'')
\tilde{\cal{G}}({\bf k}'',i\omega) \tilde{T}_{k''k}(i\omega)  .
\label{eq:kLambda}
\end{eqnarray}
Acting on both sides of (\ref{eq:GammafromLambda}), using (\ref{eq:kLambda})
to replace the left hand side by
$\hat{\bf k}'_{\ell} \tilde{\Lambda}^{\alpha}$,
and noting that by the symmetry of the scattering potential
\begin{equation}
\sum_{k} \hat{\bf k}_{\ell}
\rightarrow \hat{\bf k}'_{\ell} \sum_{k} (\hat{\bf k}' \cdot \hat{\bf k})
\label{eq:Vsymmetry}
\end{equation}
we find that
\begin{eqnarray}
\tilde{\Lambda}^{\alpha} = n_{i} \sum_{k} &&
(\hat{\bf k}' \cdot \hat{\bf k}) \tilde{\tau}_{\alpha}
\tilde{T}_{k'k}(i\omega+i\Omega)
\tilde{\cal{G}}({\bf k},i\omega + i\Omega)
\tilde{\tau}_{\alpha} \nonumber\\
&& \times \left[ \tilde{\openone} + \tilde{\Lambda}^{\alpha} \right]
\tilde{\cal{G}}({\bf k},i\omega) \tilde{T}_{kk'}(i\omega)
\label{eq:Lambdaequals}
\end{eqnarray}
where ${\bf k}'$ is the final momentum of a scattering event.

An expression for the T-matrix in terms of the scattering potential
$V_{kk'}$ and the Matsubara self-energy $\Sigma(i\omega)$
(itself a function of the scattering potential)
can be obtained by evaluating the diagram series in figure
\ref{fig:vertcorrbub}(c).
\begin{equation}
\tilde{T}_{kk'}(i\omega) = \tilde{\tau}_{3} V_{kk'} + \sum_{k_{1}}
\tilde{\tau}_{3} \tilde{\cal{G}}({\bf k_{1}},i\omega) \tilde{\tau}_{3}
V_{kk_{1}} V_{k_{1}k'} + \cdots
\label{eq:Tmatrix1}
\end{equation}
As discussed
in Appendix \ref{app:barebubble}, at temperatures low compared to the
gap maximum, quasiparticles are only generated in the small regions of
k-space surrounding each of the four gap nodes. Hence we can let
\begin{equation}
\sum_{k} \rightarrow \sum_{j=1}^{4}\int
\frac{d^{2}p}{(2\pi)^{2}v_{f}v_{2}}
\label{eq:sum2scaledint}
\end{equation}
and note that the initial and final momenta of a scattering event
must always be approximately equal to the k-space location of one of the
four nodes.  Thus if $j$ and $j'$ are node indices (1 to 4), the 
scattering potential takes the form
\begin{equation}
V_{kk'} \rightarrow V_{jj'} \rightarrow (\underline{V})_{jj'}
\label{eq:Vmatrix}
\end{equation}
where $\underline{V}$ is a 4x4 matrix in node space.  Due to
the symmetry of the nodes, $\underline{V}$ consists of only three
independent parameters: $V_{1}$ for intra-node scattering,
$V_{2}$ for adjacent-node scattering, and $V_{3}$ for opposite-node
scattering. (see Fig.\ \ref{fig:scatteringpotential} in Sec.\ \ref{sec:dos})
\begin{equation}
\underline{V} = \left(
\begin{array}{cccc}
V_{1} & V_{2} & V_{3} & V_{2} \\
V_{2} & V_{1} & V_{2} & V_{3} \\
V_{3} & V_{2} & V_{1} & V_{2} \\
V_{2} & V_{3} & V_{2} & V_{1}
\end{array} \right)
\label{eq:Vdef}
\end{equation}
Hence (\ref{eq:Tmatrix1}) becomes
\begin{eqnarray}
\tilde{T} _{jj'} && (i\omega) =
\tilde{\tau}_{3} (\underline{V})_{jj'} \nonumber \\
&& + \tilde{\tau}_{3} \left( \int\frac{d^{2}p}{(2\pi)^{2}v_{f}v_{2}}
\tilde{\cal{G}}({\bf p},i\omega) \right) \tilde{\tau}_{3}
(\underline{V}^{2})_{jj'} + \cdots
\label{eq:Tmatrix2}
\end{eqnarray}
Performing the node integration we see that the integral of the Green's
function about a node is a scalar in Nambu space (proportional to
$\tilde{\openone}$).  Since $\tilde{\tau}_{3}$ raised to an even power
is equal to $\tilde{\openone}$, the T-matrix splits into a
$\tilde{\tau}_{3}$ component and a $\tilde{\openone}$ component.
Summing the resulting geometric series we find that
\begin{eqnarray}
&& \tilde{T}_{jj'} = T^{a}_{jj'} \tilde{\tau}_{3} + T^{b}_{jj'}
\tilde{\openone} \nonumber \\
&& T^{a}_{jj'} = \left( \frac{\underline{V}}{1 - F'(i\omega)^{2}
\underline{V}^{2}} \right)_{jj'} \nonumber \\
&& T^{b}_{jj'} = \left( \frac{-F'(i\omega) \underline{V}^{2}}
{1 - F'(i\omega)^{2} \underline{V}^{2}} \right)_{jj'}
\label{eq:Tmatrix3}
\end{eqnarray}
where
\begin{equation}
F'(i\omega) = \frac{F(i\omega - \Sigma(i\omega))}{4 \pi v_{f} v_{2}}
\label{eq:F'def}
\end{equation}
\begin{equation}
F(x) \equiv x \ln \left( 1 - \frac{p_{0}^{2}}{x^{2}} \right)
\label{eq:Fdef}
\end{equation}
and $p_{0}$ is the large scaled momentum cutoff defined in Sec.\
\ref{sec:dos}.

Taking the Nambu space trace of (\ref{eq:Lambdaequals}), cyclically
permuting within the trace, and replacing the momentum sum with a
scaled integral about the nodes via (\ref{eq:sum2scaledint}) yields
\begin{eqnarray}
\mbox{Tr} \, \tilde{\Lambda}^{\alpha} &=& \mbox{Tr} \Bigl[ n_{i}
\int\frac{d^{2}p}{(2\pi)^{2}v_{f}v_{2}} \tilde{\cal{G}}({\bf p},i\omega)
\Bigr. \nonumber \\
&& \times \sum_{j=1}^{4} \left( \hat{\bf k}^{j'} \cdot \hat{\bf k}^{j}
\right) \tilde{T}_{jj'}(i\omega) \tilde{\tau}_{\alpha}
\tilde{T}_{j'j}(i\omega+i\Omega) \nonumber \\
&& \times \Bigl. \tilde{\cal{G}}({\bf p},i\omega+i\Omega)
\tilde{\tau}_{\alpha} \left( \tilde{\openone} + \tilde{\Lambda}^{\alpha}
\right) \Bigr] .
\label{eq:TrLambda1}
\end{eqnarray}
Defining node $j'$ to be node 1 we can write for $j=1,2,3,4$
\begin{eqnarray*}
{\bf k}^{j} &=& \left\{ (\pm\frac{\pi}{2},\pm\frac{\pi}{2})
\right\} \nonumber\\
\hat{\bf k}^{j'} \cdot \hat{\bf k}^{j} &=& \{ 1, 0, -1, 0 \} \nonumber\\
\tilde{T}_{jj'} = \tilde{T}_{j'j} & \equiv &
\left\{ \tilde{T}_{1}, \tilde{T}_{2}, \tilde{T}_{3}, \tilde{T}_{2} \right\}
\end{eqnarray*}
and therefore
\begin{eqnarray}
\sum_{j=1}^{4} && \left( \hat{\bf k}^{j'} \cdot \hat{\bf k}^{j} \right)
\tilde{T}_{jj'}(i\omega) \tilde{\tau}_{\alpha}
\tilde{T}_{j'j}(i\omega+i\Omega) \nonumber \\
&& = \tilde{T}_{n}(i\omega) \tilde{\tau}_{\alpha}
\tilde{T}_{n}(i\omega+i\Omega) \bigl[ |_{n=1} - |_{n=3} \bigr] \nonumber \\
&& = \frac{4\pi v_{f} v_{2}}{n_{i}} \Bigl( \gamma^{\alpha}_{A}(i\omega,i\Omega)
+ \gamma^{\alpha}_{B}(i\omega,i\Omega) \tilde{\tau}_{3} \Bigr)
\tilde{\tau}_{\alpha}
\label{eq:sumTtauT}
\end{eqnarray}
where
\begin{eqnarray}
\gamma^{\alpha}_{A} \equiv \frac{n_{i}}{4\pi v_{f}v_{2}}
&& \left( \eta_{\alpha} T^{a}_{n}(i\omega) T^{a}_{n}(i\omega+i\Omega)
\right. \nonumber \\
&& \left. + T^{b}_{n}(i\omega) T^{b}_{n}(i\omega+i\Omega) \right)
\bigl[ |_{n=1} - |_{n=3} \bigr]
\label{eq:gammaAdef}
\end{eqnarray}
\begin{eqnarray}
\gamma^{\alpha}_{B} \equiv \frac{n_{i}}{4\pi v_{f}v_{2}}
&& \left( \eta_{\alpha} T^{b}_{n}(i\omega) T^{a}_{n}(i\omega+i\Omega)
\right. \nonumber \\
&& \left. + T^{a}_{n}(i\omega) T^{b}_{n}(i\omega+i\Omega) \right)
\bigl[ |_{n=1} - |_{n=3} \bigr]
\label{eq:gammaBdef}
\end{eqnarray}
and we define
\begin{equation}
\eta_{\alpha} \equiv \left\{ \begin{array}{ll}
+1 & \mbox{for $\alpha = 0,3$} \\
-1 & \mbox{for $\alpha = 1,2$} \end{array} \right. .
\label{eq:etadef}
\end{equation}
Plugging (\ref{eq:sumTtauT}) into (\ref{eq:TrLambda1}) and defining
\begin{equation}
\tilde{I}^{\alpha}(i\omega,i\Omega) = \int \frac{d^{2}p}{\pi}
\tilde{\cal{G}}({\bf p},i\omega) \tilde{\tau}_{\alpha}
\tilde{\cal{G}}({\bf p},i\omega + i\Omega) \tilde{\tau}_{\alpha}
\label{eq:Idef}
\end{equation}
and
\begin{equation}
\tilde{I}^{\alpha}_{3}(i\omega,i\Omega) = \int \frac{d^{2}p}{\pi}
\tilde{\cal{G}}({\bf p},i\omega)
\tilde{\tau}_{3} \tilde{\tau}_{\alpha}
\tilde{\cal{G}}({\bf p},i\omega + i\Omega) \tilde{\tau}_{\alpha}
\label{eq:I3def}
\end{equation}
we find that
\begin{equation}
\mbox{Tr} \, \tilde{\Lambda}^{\alpha} = \mbox{Tr} \,
\left[ \gamma^{\alpha}_{A} \tilde{I}^{\alpha} (\tilde{\openone}
+ \tilde{\Lambda}^{\alpha}) \right] + \mbox{Tr} \,
\left[ \gamma^{\alpha}_{B} \tilde{I}^{\alpha}_{3} (\tilde{\openone}
+ \tilde{\Lambda}^{\alpha}) \right] .
\label{eq:TrLambda2}
\end{equation}
Similarly, multiplying (\ref{eq:Lambdaequals}) by $\tilde{\tau}_{3}$
and repeating our steps we find that
\begin{equation}
\mbox{Tr} \, \left[ \tilde{\tau}_{3} \tilde{\Lambda}^{\alpha} \right]
= \mbox{Tr} \,
\left[ \gamma^{\alpha}_{B} \tilde{I}^{\alpha} (\tilde{\openone}
+ \tilde{\Lambda}^{\alpha}) \right] + \mbox{Tr} \,
\left[ \gamma^{\alpha}_{A} \tilde{I}^{\alpha}_{3} (\tilde{\openone}
+ \tilde{\Lambda}^{\alpha}) \right] .
\label{eq:Trtau3Lambda}
\end{equation}

At this point it is useful to carry out the momentum integrals in 
(\ref{eq:Idef},\ref{eq:I3def}) to obtain an explicit form for
$\tilde{I}^{\alpha}$ and $\tilde{I}^{\alpha}_{3}$.
Recalling the form of the Matsubara Green's function from
Sec.\ \ref{sec:dos}, noting that $p_{1} = \varepsilon_{k}$
and $p_{2} = \Delta_{k}$, and defining
\begin{mathletters}
\label{eq:f12def}
\begin{eqnarray}
f_{1} & \equiv & i\omega - \Sigma(i\omega) \\
f_{2} & \equiv & i\omega + i\Omega - \Sigma(i\omega + i\Omega)
\end{eqnarray}
\end{mathletters}
we can write
\begin{equation}
\tilde{\cal{G}}({\bf p},i\omega) = \frac{f_{1} \tilde{\openone}
+ p_{1} \tilde{\tau}_{3} + p_{2} \tilde{\tau}_{1}}{f_{1}^{2} - p^{2}} .
\label{eq:Greenpf1}
\end{equation}
Further, using the definition of $\eta_{\alpha}$ from (\ref{eq:etadef})
and similarly defining
\begin{equation}
\eta'_{\alpha} \equiv \left\{ \begin{array}{ll}
+1 & \mbox{for $\alpha = 0,1$} \\
-1 & \mbox{for $\alpha = 2,3$} \end{array} \right.
\label{eq:etaprime}
\end{equation}
we find that
\begin{equation}
\tilde{\tau}_{\alpha} \tilde{\cal{G}}({\bf p},i\omega + i\Omega)
\tilde{\tau}_{\alpha} = \frac{f_{2} \tilde{\openone}
+ \eta_{\alpha} p_{1} \tilde{\tau}_{3}
+ \eta'_{\alpha} p_{2} \tilde{\tau}_{1}}{f_{2}^{2} - p^{2}} .
\label{eq:Greenpf2}
\end{equation}
Thus, plugging (\ref{eq:Greenpf1}) and (\ref{eq:Greenpf2}) into
(\ref{eq:Idef}), noting that $p_{1} = p \cos\theta$ and
$p_{2} = p \sin\theta$, and performing the angular integral
we find that
\begin{equation}
\tilde{I}^{\alpha} = I^{\alpha} \tilde{\openone}
= \tilde{\openone} \int_{0}^{p_{0}} 2p \frac{f_{1}f_{2}
+ a_{\alpha} p^{2}}{(f_{1}^{2} - p^{2})(f_{2}^{2} - p^{2})} dp
\label{eq:Ipreint}
\end{equation}
where
\begin{equation}
a_{\alpha} \equiv \frac{\eta_{\alpha} + \eta'_{\alpha}}{2}
= \left\{ \begin{array}{lll}
+1 & \mbox{for $\alpha = 0$} \\
\ \ 0 & \mbox{for $\alpha = 1,3$} \\
-1 & \mbox{for $\alpha = 2$} \end{array} \right. .
\label{eq:adef}
\end{equation}
Factoring the integrand and performing the $p$-integral yields
\begin{equation}
I^{\alpha}(i\omega,i\Omega) = \frac{(f_{1} + a_{\alpha} f_{2}) F(f_{2})
- (f_{2} + a_{\alpha} f_{1}) F(f_{1})}{f_{2}^{2} - f_{1}^{2}}
\label{eq:If1f2}
\end{equation}
where $F(x)$ is defined via (\ref{eq:Fdef}).
Similarly, acting on (\ref{eq:I3def}) yields that
\begin{equation}
\tilde{I}^{\alpha}_{3}
= \tilde{\tau}_{3} \int_{0}^{p_{0}} 2p \frac{f_{1}f_{2}
+ a'_{\alpha} p^{2}}{(f_{1}^{2} - p^{2})(f_{2}^{2} - p^{2})} dp
\label{eq:I3preint}
\end{equation}
where
\begin{equation}
a'_{\alpha} \equiv \frac{\eta_{\alpha} - \eta'_{\alpha}}{2}
= \left\{ \begin{array}{lll}
+1 & \mbox{for $\alpha = 3$} \\
\ \ 0 & \mbox{for $\alpha = 0,2$} \\
-1 & \mbox{for $\alpha = 1$} \end{array} \right. .
\label{eq:a'def}
\end{equation}
It is easy to see that $a'_{\alpha} = a_{\alpha + 1}$ (where the
index addition is defined modulo 4).  Hence
\begin{mathletters}
\label{eq:IandI3}
\begin{eqnarray}
\tilde{I}^{\alpha}(i\omega,i\Omega) &=& I^{\alpha} \tilde{\openone} \\
\tilde{I}^{\alpha}_{3}(i\omega,i\Omega) &=& I^{\alpha + 1}
\tilde{\tau}_{3}
\end{eqnarray}
\end{mathletters}
where $I^{\alpha}$ is given by (\ref{eq:If1f2}).

Now that $\tilde{I}^{\alpha}$ and $\tilde{I}^{\alpha}_{3}$ have
been evaluated, (\ref{eq:TrLambda2}) and (\ref{eq:Trtau3Lambda})
become a set of coupled equations for
$\mbox{Tr} \, \tilde{\Lambda}^{\alpha}$ and
$\mbox{Tr} \, [\tilde{\tau}_{3} \tilde{\Lambda}^{\alpha}]$:
\begin{mathletters}
\label{eq:Lambdacoupled}
\begin{eqnarray}
\mbox{Tr} \, \tilde{\Lambda}^{\alpha} = \gamma^{\alpha}_{A} I^{\alpha}
(2 + \mbox{Tr} \, \tilde{\Lambda}^{\alpha}) + \gamma^{\alpha}_{B}
I^{\alpha + 1} \mbox{Tr} [\tilde{\tau}_{3}
\tilde{\Lambda}^{\alpha}] \\
\mbox{Tr} [\tilde{\tau}_{3} \tilde{\Lambda}^{\alpha}] =
\gamma^{\alpha}_{B} I^{\alpha}
(2 + \mbox{Tr} \, \tilde{\Lambda}^{\alpha}) + \gamma^{\alpha}_{A}
I^{\alpha + 1} \mbox{Tr} [\tilde{\tau}_{3}
\tilde{\Lambda}^{\alpha}]
\end{eqnarray}
\end{mathletters}
Solving simultaneously yields
\begin{equation}
\mbox{Tr} \left[ \tilde{\openone} + \tilde{\Lambda}^{\alpha} \right]
= \frac{2}{1 - \gamma^{\alpha}_{A} I^{\alpha} \left( 1 +
\frac{\gamma^{\alpha}_{B}}{\gamma^{\alpha}_{A}} \frac{\gamma^{\alpha}_{B}
I^{\alpha + 1}}{1 - \gamma^{\alpha}_{A} I^{\alpha + 1}} \right)} .
\label{eq:Tr1+L}
\end{equation}
This is a very useful result since using (\ref{eq:GammafromLambda}),
(\ref{eq:sum2scaledint}), (\ref{eq:Idef}),
and (\ref{eq:IandI3}) with (\ref{eq:polvert1}) yields that
\begin{eqnarray}
\tensor{\Pi}^{g\ell\alpha}(i\Omega) = && \frac{1}{4\pi v_{f} v_{2}}
\sum_{j=1}^{4} {\bf v}_{\ell}^{(j)} {\bf v}_{\ell}^{(j)} \nonumber\\
&& \times \frac{1}{\beta} \sum_{i\omega} g^{2} I^{\alpha}
\mbox{Tr} \left[ \tilde{\openone} + \tilde{\Lambda}^{\alpha} \right] .
\label{eq:polvert2}
\end{eqnarray}
Thus plugging (\ref{eq:Tr1+L}) into (\ref{eq:polvert2}) and making
use of (\ref{eq:vvsum}) from Appendix \ref{app:barebubble} we find that
\begin{equation}
\tensor{\Pi}^{g\ell\alpha} = \Pi^{g\ell\alpha} \tensor{\openone}
\end{equation}
where
\begin{equation}
\Pi^{g\ell\alpha}(i\Omega) = \frac{v_{\ell}^{2}}{\pi v_{f} v_{2}}
\frac{1}{\beta} \sum_{i\omega} g^{2}
J^{\alpha}(i\omega,i\Omega)
\label{eq:polvert3}
\end{equation}
\begin{equation}
J^{\alpha} \equiv
\frac{I^{\alpha}}{1 - \gamma^{\alpha}_{A} I^{\alpha} \left( 1 +
\frac{\gamma^{\alpha}_{B}}{\gamma^{\alpha}_{A}}
\frac{\gamma^{\alpha}_{B} I^{\alpha + 1}}{1 - \gamma^{\alpha}_{A}
I^{\alpha + 1}} \right)}
\label{eq:Jdef}
\end{equation}
and we note that $I^{\alpha}$, $I^{\alpha + 1}$, $\gamma^{\alpha}_{A}$,
and $\gamma^{\alpha}_{B}$ are all functions of $i\omega$ and $i\Omega$.

Provided the input self-energy is of a proper functional form,
$J^{\alpha}(z,i\Omega)$
will be analytic throughout the complex plane except for two branch
cuts at $\mbox{Im}\, z = 0$ and $\mbox{Im}\, z = -\Omega$.  Thus,
evaluating the Matsubara sum \cite{mah90} we pick up a contribution from
each of the branch cuts of the summand.  Consequently, it is useful to
consider the form of $J^{\alpha}(z,i\Omega)$ above and below each of
the branch cuts.  Upon examination of the frequency dependence of this
function via (\ref{eq:Jdef},\ref{eq:gammaAdef}-\ref{eq:I3def}) it is
clear that the internal and external frequencies, $i\omega$ and
$i\Omega$, enter only through functional couplets of the form
\begin{equation}
P(i\omega,i\Omega) = A(i\omega) B(i\omega + i\Omega) .
\label{eq:couplet}
\end{equation}
Furthermore, due to the defined analytic structure of Matsubara
Green's functions, the functions composing these couplets always
have a Matsubara-like analytic structure and satisfy
\begin{eqnarray}
A(i\omega \rightarrow \omega + i\delta) &=& A_{ret}(\omega) \\
A(i\omega \rightarrow \omega - i\delta) &=& A_{ret}^{*}(\omega) .
\label{eq:MatStructure}
\end{eqnarray}
Consider the form of such a couplet above and below the branch cuts of
our summand.  Defining
\begin{mathletters}
\label{eq:P1234def}
\begin{eqnarray}
P_{1}(\omega,i\Omega) & \equiv &
P(\omega + i\delta,i\Omega) \\
P_{2}(\omega,i\Omega) & \equiv &
P(\omega - i\delta,i\Omega) \\
P_{3}(\omega,i\Omega) & \equiv &
P(\omega - i\Omega + i\delta,i\Omega) \\
P_{4}(\omega,i\Omega) & \equiv &
P(\omega - i\Omega - i\delta,i\Omega)
\end{eqnarray}
\end{mathletters}
and continuing $i\Omega \rightarrow \Omega + i\delta$ we see that
\begin{mathletters}
\label{eq:Prelations}
\begin{eqnarray}
P_{3}(\omega,\Omega) &=&
P_{2}(\omega - \Omega,\Omega) \\
P_{4}(\omega,\Omega) &=&
P_{1}^{*}(\omega - \Omega,\Omega) .
\end{eqnarray}
\end{mathletters}
Since $J^{\alpha}$ is composed of such couplets, if we similarly define
\begin{mathletters}
\label{eq:J1234def}
\begin{eqnarray}
J^{\alpha}_{1}(\omega,i\Omega) & \equiv &
J^{\alpha}(\omega + i\delta,i\Omega) \\
J^{\alpha}_{2}(\omega,i\Omega) & \equiv &
J^{\alpha}(\omega - i\delta,i\Omega) \\
J^{\alpha}_{3}(\omega,i\Omega) & \equiv &
J^{\alpha}(\omega - i\Omega + i\delta,i\Omega) \\
J^{\alpha}_{4}(\omega,i\Omega) & \equiv &
J^{\alpha}(\omega - i\Omega - i\delta,i\Omega)
\end{eqnarray}
\end{mathletters}
it follows that
\begin{mathletters}
\label{eq:Jrelations}
\begin{eqnarray}
J^{\alpha}_{3}(\omega,\Omega) &=&
J^{\alpha}_{2}(\omega - \Omega,\Omega) \\
J^{\alpha}_{4}(\omega,\Omega) &=&
\bigl. J^{\alpha}_{1} \bigr.^{*}(\omega - \Omega,\Omega) .
\end{eqnarray}
\end{mathletters}
These relations will be very helpful in what follows.
Now we can proceed with the Matsubara sum.
As in Appendix \ref{app:barebubble}, it is best to treat
the frequency-independent coupling and frequency-dependent coupling
cases separately.

For $g = \{ e,s \}$ (frequency-independent coupling), the sum is
straightforward.  Adding the contributions of the two branch cuts
and using the definitions in (\ref{eq:J1234def}) yields
\begin{eqnarray}
\Pi^{g\ell\alpha}(i\Omega) =  && i \frac{g^{2}}{2 \pi^{2}}
\frac{v_{\ell}^{2}}{v_{f} v_{2}} \int_{-\infty}^{\infty}
 d\omega \, n_{F}(\omega) \nonumber\\
&& \times \Bigl[ J^{\alpha}_{1}(\omega,i\Omega)
- J^{\alpha}_{2}(\omega,i\Omega) \Bigr. \nonumber\\
&& \Bigl. + J^{\alpha}_{3}(\omega,i\Omega)
- J^{\alpha}_{4}(\omega,i\Omega) \Bigr]
\label{eq:polvertindep}
\end{eqnarray}
where $n_{F}(\omega)$ is the Fermi function.

For $g = \omega + \Omega/2$ (frequency-dependent coupling), noting
the technical issues discussed in the analogous stage of the bare
bubble calculation (see Appendix \ref{app:barebubble}), we can
proceed as above.  Adding the contributions of the two branch cuts
and using (\ref{eq:J1234def}) we find that
\begin{eqnarray}
\Pi^{g\ell\alpha}(i\Omega) && = i \frac{1}{2 \pi^{2}}
\frac{v_{\ell}^{2}}{v_{f} v_{2}} \int_{-\infty}^{\infty}
d\omega \, n_{F}(\omega) \nonumber\\
&& \times \Bigl[ \Bigl( \omega + \frac{i\Omega}{2} \Bigr)^{2}
\left( J^{\alpha}_{1}(\omega,i\Omega)
- J^{\alpha}_{2}(\omega,i\Omega) \right) \Bigr. \nonumber\\
&& \Bigl. + \Bigl( \omega - \frac{i\Omega}{2} \Bigr)^{2}
\left( J^{\alpha}_{3}(\omega,i\Omega)
- J^{\alpha}_{4}(\omega,i\Omega) \right) \Bigr] .
\label{eq:polvertdep}
\end{eqnarray}

In either case, continuing $i\Omega \rightarrow \Omega + i\delta$,
making use of the relations in (\ref{eq:Jrelations}), and shifting
$\omega \rightarrow \omega + \Omega$ in the last two terms, we obtain
the retarded polarization function
\begin{eqnarray}
\Pi^{g\ell\alpha}_{ret}(\Omega) && = i \frac{1}{2 \pi^{2}}
\frac{v_{\ell}^{2}}{v_{f} v_{2}} \int_{-\infty}^{\infty}
d\omega \, g^{2} \nonumber\\
&& \times \Bigl[ n_{F}(\omega)
\bigl( J^{\alpha}_{1}(\omega,\Omega)
- J^{\alpha}_{2}(\omega,\Omega) \bigr) \Bigr. \nonumber\\
&& \Bigl. -\ n_{F}(\omega + \Omega)
\bigl( \bigl. J^{\alpha}_{1} \bigr.^{*}(\omega,\Omega)
- J^{\alpha}_{2}(\omega,\Omega) \bigr) \Bigr]
\label{eq:retpolvert}
\end{eqnarray}
which is valid for all three coupling parameters.  Taking the
imaginary part, noting that $\mbox{Re} [z] = \mbox{Re} [z^{*}]$,
and expanding $J^{\alpha}_{1}$ and $J^{\alpha}_{2}$ yields
\begin{eqnarray}
\mbox{Im}\, && \Pi^{g\ell\alpha}_{ret}(\Omega) = \frac{1}{2 \pi^{2}}
\frac{v_{\ell}^{2}}{v_{f} v_{2}} \int_{-\infty}^{\infty} d\omega \, g^{2}
\left( n_{F}(\omega + \Omega) - n_{F}(\omega) \right) \nonumber\\
&& \times \mbox{Re} \left[ \frac{I^{\alpha}_{2}}{1 - \gamma^{\alpha}_{A2}
I^{\alpha}_{2} \left( 1 + \frac{\gamma^{\alpha}_{B2}}{\gamma^{\alpha}_{A2}}
\frac{\gamma^{\alpha}_{B2} I^{\alpha + 1}_{2}}{1 - \gamma^{\alpha}_{A2}
I^{\alpha + 1}_{2}} \right) } \right. \nonumber\\
&& \;\;\;\;\;\;\: - \left. \frac{I^{\alpha}_{1}}{1 - \gamma^{\alpha}_{A1}
I^{\alpha}_{1} \left( 1 + \frac{\gamma^{\alpha}_{B1}}{\gamma^{\alpha}_{A1}}
\frac{\gamma^{\alpha}_{B1} I^{\alpha + 1}_{1}}{1 - \gamma^{\alpha}_{A1}
I^{\alpha + 1}_{1}} \right) } \right]
\label{eq:imretpolvert}
\end{eqnarray}
where
\begin{equation}
I^{\alpha}_{1}(\omega,\Omega) = \frac{(f_{1} \! + \! a_{\alpha} f_{2})
F(f_{2}) - (f_{2} \! + \! a_{\alpha} f_{1}) F(f_{1})}
{f_{2}^{2} - f_{1}^{2}}
\label{eq:I1def}
\end{equation}
\begin{equation}
I^{\alpha}_{2}(\omega,\Omega) = \frac{(f_{1}^{*} \! + \! a_{\alpha} f_{2})
F(f_{2}) - (f_{2} \! + \! a_{\alpha} f_{1}^{*}) F(f_{1}^{*})}
{f_{2}^{2} - \left. f_{1}^{*} \right.^{2}}
\label{eq:I2def}
\end{equation}
\begin{mathletters}
\label{eq:f12retdef}
\begin{eqnarray}
f_{1} &=& \omega - \Sigma_{ret}(\omega) \\
f_{2} &=& \omega + \Omega - \Sigma_{ret}(\omega + \Omega) .
\end{eqnarray}
\end{mathletters}
\begin{eqnarray}
\gamma^{\alpha}_{A1} \equiv \frac{n_{i}}{4\pi v_{f}v_{2}}
&& \left( \eta_{\alpha} T^{a}_{n}(\omega) T^{a}_{n}(\omega+\Omega)
\right. \nonumber \\
&& \left. + T^{b}_{n}(\omega) T^{b}_{n}(\omega+\Omega) \right)
\bigl[ |_{n=1} - |_{n=3} \bigr]
\label{eq:gammaA1def}
\end{eqnarray}
\begin{eqnarray}
\gamma^{\alpha}_{A2} \equiv \frac{n_{i}}{4\pi v_{f}v_{2}}
&& \left( \eta_{\alpha} T^{a}_{n}(\omega)^{*}\, T^{a}_{n}(\omega+\Omega)
\right. \nonumber \\
&& \left. + T^{b}_{n}(\omega)^{*}\, T^{b}_{n}(\omega+\Omega) \right)
\bigl[ |_{n=1} - |_{n=3} \bigr]
\label{eq:gammaA2def}
\end{eqnarray}
\begin{eqnarray}
\gamma^{\alpha}_{B1} \equiv \frac{n_{i}}{4\pi v_{f}v_{2}}
&& \left( \eta_{\alpha} T^{b}_{n}(\omega) T^{a}_{n}(\omega+\Omega)
\right. \nonumber \\
&& \left. + T^{a}_{n}(\omega) T^{b}_{n}(\omega+\Omega) \right)
\bigl[ |_{n=1} - |_{n=3} \bigr]
\label{eq:gammaB1def}
\end{eqnarray}
\begin{eqnarray}
\gamma^{\alpha}_{B2} \equiv \frac{n_{i}}{4\pi v_{f}v_{2}}
&& \left( \eta_{\alpha} T^{b}_{n}(\omega)^{*}\, T^{a}_{n}(\omega+\Omega)
\right. \nonumber \\
&& \left. + T^{a}_{n}(\omega)^{*}\, T^{b}_{n}(\omega+\Omega) \right)
\bigl[ |_{n=1} - |_{n=3} \bigr]
\label{eq:gammaB2def}
\end{eqnarray}
\begin{equation}
T^{a}_{n}(\omega) = \left( \frac{\underline{V}}{1 - F'(\omega)^{2}
\underline{V}^{2}} \right)_{n1}
\label{eq:Tmata}
\end{equation}
\begin{equation}
T^{b}_{n}(\omega) = \left( \frac{-F'(\omega) \underline{V}^{2}}
{1 - F'(\omega)^{2} \underline{V}^{2}} \right)_{n1}
\label{eq:Tmatb}
\end{equation}
\begin{equation}
F'(\omega) = \frac{F(\omega - \Sigma_{ret}(\omega))}{4 \pi v_{f} v_{2}}
\label{eq:F'retdef}
\end{equation}
and $F(x)$ is defined in (\ref{eq:Fdef}).
The above equations define the imaginary part of the generalized
retarded polarization function including ladder corrections to the
vertex.  By specifying different input parameters we can use it
to obtain the electrical, thermal, and spin conductivity.

\section{Numerical Analysis of Universal Limit Vertex Corrections}
\label{app:numerical}
In Secs.\ \ref{ssec:elecvert}, \ref{ssec:thermvert}, and \ref{ssec:spinvert},
the vertex correction factors for the universal limit electrical and
thermal/spin conductivities were found to be
\begin{eqnarray}
&& \beta_{vc} = \nonumber \\
&& \frac{1 + 2 \left( \gamma^{(0)}_{A2} - \gamma^{(0)}_{A1}
+\frac{\left. \gamma^{(0)}_{B1} \right.^{2}}{1-\gamma^{(0)}_{A1}} \right)
\ln\frac{p_{0}}{\Gamma_{0}} \left( \ln\frac{p_{0}}{\Gamma_{0}} - 1 \right)}
{\left( 1 - 2\gamma^{(0)}_{A2}
\ln\frac{p_{0}}{\Gamma_{0}} \right) \left( 1 - 2\left( \gamma^{(0)}_{A1} -
\frac{\left. \gamma^{(0)}_{B1} \right.^{2}}{1-\gamma^{(0)}_{A1}} \right)
\left( \ln\frac{p_{0}}{\Gamma_{0}} - 1 \right) \right) } \nonumber \\ &&
\label{eq:appbetaVC}
\end{eqnarray}
\begin{equation}
\beta^{T,s}_{vc} = \frac{\frac{1}{2}}{1 - \gamma^{(0)}_{A2}}
+ \frac{\frac{1}{2}}{1 + \gamma^{(0)}_{A1} \left( 1 +
\frac{\gamma^{(0)}_{B1}}{\gamma^{(0)}_{A1}}
\frac{\gamma^{(0)}_{B1} \left( 2 \ln \frac{p_{0}}{\Gamma_{0}} - 2 \right)}
{1 - \gamma^{(0)}_{A1} \left( 2 \ln \frac{p_{0}}{\Gamma_{0}} - 2 \right)}
\right)}
\label{eq:appbetaVCthermal}
\end{equation}
where the $\gamma$'s and their constituent functions are given in
(\ref{eq:F'0}-\ref{eq:gammas0}).  Here we shall numerically compute
both factors as a function of impurity density and scattering potential.
To facilitate the computation, it is convenient
to make all quantities dimensionless by expressing energies in units
of $p_{0}$ and lengths in units of $2a$.  This choice of units sets the
frequently encountered constant, $4\pi v_{f}v_{2}$, equal to 1.
For a particular set of input parameters, the computation is done in two
steps.  First, solve self-consistently for the zero-frequency scattering
rate, $\Gamma_{0}$, via
\begin{equation}
\Sigma_{ret}(\omega)=n_{i} T_{11}^{b}(\omega)
\label{eq:SigEQnT}
\end{equation}
which (in the universal limit with our choice of units) reduces to
\begin{equation}
2 n_{i} \ln\frac{1}{\Gamma_{0}} \left( \frac{\underline{V}^{2}}
{1+ \left( 2\Gamma_{0}\ln\frac{1}{\Gamma_{0}} \right)^{2} \underline{V}^{2}}
\right)_{1,1} = 1 .
\label{eq:selfconGam0}
\end{equation}
Then given $\Gamma_{0}$, plug into (\ref{eq:appbetaVC}) and
(\ref{eq:appbetaVCthermal}) to obtain the vertex corrections.

The result will depend on the set of four input parameters which determine
the impurity density and scattering potential, $\{n_{i},V_{1},V_{2},V_{3}\}$.
In our units, $n_{i}=4z$ where $z$ is the substitutional fraction of impurities.
Furthermore, it is convenient to parametrize the scattering potential via
\begin{equation}
V_{1} = V_{scale} \tan \left( \frac{\pi}{2} d \right) \;\;\;\;\;\;
R_{2} = \frac{V_{2}}{V_{1}} \;\;\;\;\;\; R_{3} = \frac{V_{3}}{V_{1}}
\label{eq:vtand}
\end{equation}
where $V_{scale}$ is a dimensionless constant and $d$ ranges
from 0 to 1.  Hence, the scattering anisotropy is given by $R_{2}$ and $R_{3}$
while the scattering strength is given by $d$.
Note that while is tempting to think of $\frac{\pi}{2}d$ as
some sort of scattering phase shift, to do so would be stretching an analogy
beyond its realm of usefulness.  The above is merely a helpful parametrization,
the aim of which
is to allow us to go smoothly from the Born limit to the unitary limit as $d$
varies from 0 to 1.  For our purposes, these limits are determined by the
denominator in (\ref{eq:selfconGam0}):
\begin{eqnarray*}
\left( 2 \Gamma_{0} \ln\frac{1}{\Gamma_{0}} \right) V_{1} \ll 1
\;\;\;\; &\rightarrow& \;\;\;\; \mbox{Born limit} \\
\left( 2 \Gamma_{0} \ln\frac{1}{\Gamma_{0}} \right) V_{1} \gg 1
\;\;\;\; &\rightarrow& \;\;\;\; \mbox{Unitary limit}
\end{eqnarray*}
Thus, for a particular range of $z$, we shall set $V_{scale}$ to a value that
allows us to evenly sample the transition from one limit to the other.
In the end, these manipulations yield a new set of four dimensionless input
parameters, $\{z,d,R_{2},R_{3}\}$.

\begin{figure}[tb]
\centerline{\psfig{file=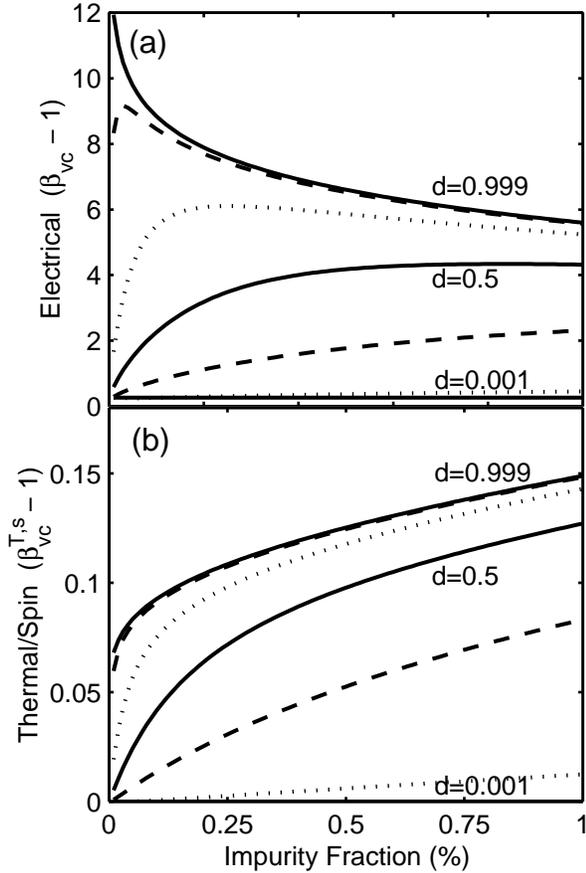}}
\vspace{0.5 cm}
\caption{Numerically calculated (a) electrical and (b) thermal/spin vertex
corrections plotted as a function of impurity fraction
($z=0.01\% \rightarrow 1\% $) for scattering strengths parametrized via
(from bottom to top) $d= \{ 0.001 \mbox{\ (Born)},$ $0.1, 0.3, 0.5, 0.7, 0.9,$
$0.999 \mbox{\ (unitary)} \}$.
In all cases we have set $V_{scale}=20$ and assumed that $V({\bf k})$
falls off slowly with $|{\bf k}|$ ($R_{2}=0.9$, $R_{3}=0.8$).}
\label{fig:VCplots1}
\end{figure}

\begin{figure}[tb]
\centerline{\psfig{file=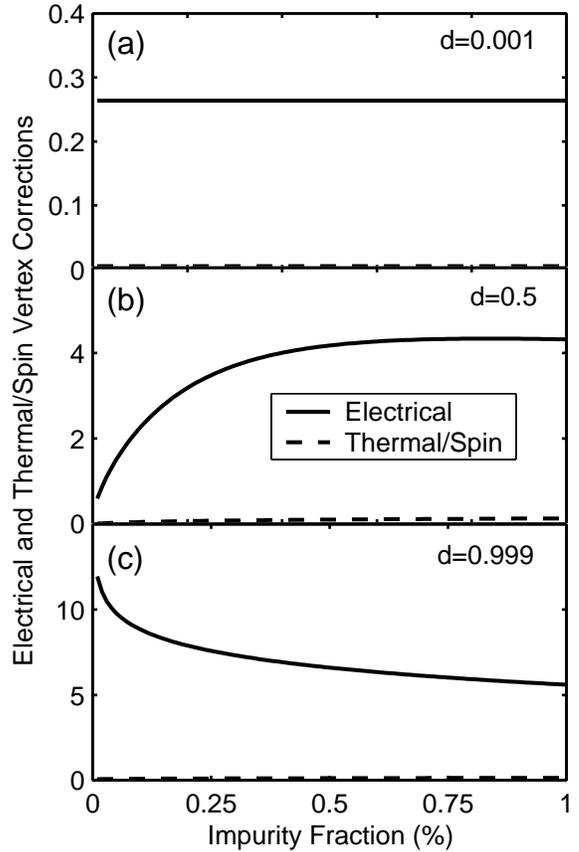}}
\vspace{0.5 cm}
\caption{Direct comparison of electrical ($\beta_{vc}-1$)
and thermal/spin ($\beta_{vc}^{T,s}-1$) vertex corrections for
(a) $d=0.001$, (b) $d=0.5$, and (c) $d=0.999$.  Note that in all
cases, the thermal/spin correction is negligible on the scale of
of the electrical correction.}
\label{fig:VCplots2}
\end{figure}

The electrical and thermal/spin vertex corrections obtained for 
a typical set of input parameters, via the procedure described above,
are plotted in Fig.\ \ref{fig:VCplots1}.  Here we have assumed that
$V({\bf k})$ falls off slowly with increasing ${\bf k}$
($R_{2}=0.9$, $R_{3}=0.8$) and have plotted the vertex corrections versus
impurity fraction ($z=0.01\% \rightarrow 1\%$) for seven scattering strengths
from Born ($d=0.001$) to unitary ($d=0.999$).  Note that the electrical
correction, $\beta_{vc}-1$, can be quite significant while the thermal/spin
correction, $\beta^{T,s}_{vc}-1$, is much smaller and vanishes as
$z\rightarrow0$.  The difference between the two cases is seen most clearly
in Fig.\ \ref{fig:VCplots2} where, for $d=0.001,0.5,0.999$, we have
replotted the electrical and thermal/spin correction factors on the same
scale.  On the scale of $\beta_{vc}-1$, it is difficult to distinguish
$\beta^{T,s}_{vc}-1$ from the x-axis.  Thus, we see that for all scattering
strengths, the thermal/spin vertex correction is negligible compared to the
electrical vertex correction.

Additional insight is gained through consideration of the intermediate stages
of the calculation which reveal that
\begin{equation}
\gamma^{(0)}_{A1} , \gamma^{(0)}_{A2} , \gamma^{(0)}_{B1}
\stackrel{<}{\sim}{\cal O}
\left( \frac{1}{\ln\frac{p_{0}}{\Gamma_{0}}} \right) .
\label{eq:smallgammas}
\end{equation}
In the Born limit, $1/\ln(p_{0}/\Gamma_{0}) \sim z$, which can truly be neglected
for the small impurity fractions of interest.  In the unitary limit,
$1/\ln(p_{0}/\Gamma_{0}) \sim 1/\ln(1/z)$, which vanishes much more slowly
but is still small compared to terms of order unity.  Thus, since we can
treat the $\gamma$'s as small quantities, the (mathematical) difference between
the electrical and thermal/spin cases is due to the manner in which the
$\gamma$'s enter (\ref{eq:appbetaVC}) and (\ref{eq:appbetaVCthermal}).
The $\gamma$'s enter $\beta_{vc}$ within order unity combinations of the form
$\gamma\ln(p_{0}/\Gamma_{0})$ (and for some parameter values even
$\gamma\ln^{2}(p_{0}/\Gamma_{0})$) which cannot be neglected.  In contrast,
the $\gamma$'s enter $\beta^{T,s}_{vc}$ on their own (that is, in direct
competition with terms of order unity) and can therefore be neglected for small
$z$.  Hence we say that electrical vertex corrections contribute even to
zeroth order in the impurity density while thermal/spin vertex corrections do not.

\section{Fermi Liquid Corrections}
\label{app:fermiliquid}
In addition to the effects of vertex corrections, transport coefficients
may be further modified due to underlying Fermi liquid interactions
between Landau quasiparticles (referred to hereafter as ``electrons'' to
avoid any confusion with the Bogoliubov quasiparticle excitations of the
superconductor).  A detailed theory of the superfluid
Fermi liquid has been developed in the literature.
\cite{mil98,xu95,leg65,lar63,lar64}
In particular, a widely applicable phenomenological approach has been
devised by Leggett. \cite{leg65}  Within this approach, we consider three
layers of Fermi liquid effects.  (1) Mass renormalization: By virtue of
our acknowledgment that ``electrons'' are really Landau quasiparticles, all
masses should be viewed as effective masses, $m^{*}$.  (2) Current
renormalization: Due to Fermi liquid interactions, the presence of a
current yields an additional drag current resulting in an overall current
renormalization.  (3) Response function modification: Fermi liquid
interactions induce an effective ``molecular field'' which couples to the
current-producing perturbation and modifies the response function,
$K_{0}(T)$, via
\begin{equation}
K_{0}(T) \rightarrow K(T) = \frac{K_{0}(T)}{1 + \lambda K_{0}(T)}
\label{eq:modrespfunc}
\end{equation}
where $\lambda$ is a constant which depends on the interaction. \cite{leg65}
In the case at hand, (1) has been built into the parameters of our
model, (2) yields a non-trivial effect which shall be considered below,
and (3) can be neglected in the $T\rightarrow0$ limit with which we are
concerned.  This last statement follows because at low $T$, few
quasiparticles are generated, response functions (such as the normal
fluid density or conductivity) are small, and the higher order correction
terms in (\ref{eq:modrespfunc}) are negligible.  Thus, the dominant
corrections to transport coefficients due to Fermi-liquid interactions
enter simply via renormalization of the various current density operators.
To determine the nature of this current renormalization,  we proceed as
follows. \cite{pin66}

In the absence of Fermi liquid interactions, a generic (bare) current density
operator takes the form
\begin{equation}
{\bf j}_{0} = \sum_{k\alpha} g_{k\alpha} {\bf v}_{k}\, \delta n_{k\alpha}
\label{eq:genbarecurrent}
\end{equation}
where ${\bf v}_{k}$ is a velocity, $g_{k\alpha}$ is a coupling
parameter (i.e.\ charge, spin, or energy), and
\begin{equation}
\delta n_{k\alpha} =  n_{k\alpha} - n_{0}(\epsilon_{k},\Delta_{k})
\label{eq:barechange}
\end{equation}
is the difference between the true electron distribution in the presence of
the current-inducing perturbation, $n_{k\alpha}$, and the (bare) equilibrium
distribution, $n_{0}(\epsilon_{k},\Delta_{k})$.  (The {\em bare} designations
refer to our neglect of Fermi liquid interactions aside from the extent to
which they are included in the velocity via mass renormalization.)

Once Fermi liquid interactions are turned on, excited electrons interact
via the Landau interaction energy, $f^{\alpha\alpha'}_{kk'}$.  In the
presence of this interaction, the electron dispersion must be modified
(or dressed) to account for the additional energy cost of
interacting with other excited electrons.
\begin{equation}
\tilde{\epsilon}_{k} = \epsilon_{k} + \sum_{k'\alpha'}
f^{\alpha\alpha'}_{kk'} \delta n_{k'\alpha'}
\label{eq:dresseps}
\end{equation}
(Note that in this context, the tilde denotes a dressed quantity, not a
Nambu matrix).  It is important to realize that the dispersion of
{\em every} electron (not just the excited ones) is dressed as long as
any electrons are excited.  Although only the excited electrons
interact, the dispersion measures the energy required to excite an
electron and must therefore account for the interactions an electron
{\em would} experience if it were excited.
Also note that although $\Delta_{k}$ should also be modified due to Fermi
Liquid interactions, we expect this effect to be less significant and
shall therefore assume for the sake of simplicity that the gap is unaffected.
Once the dispersion is dressed, we can Taylor expand in the change in
$\epsilon_{k}$ to obtain a dressed equilibrium distribution function
\begin{equation}
n_{0}(\tilde{\epsilon}_{k},\Delta_{k}) = n_{0}(\epsilon_{k},\Delta_{k})
+ \frac{\partial n_{0}}{\partial \epsilon_{k}} \sum_{k'\alpha'}
f^{\alpha\alpha'}_{kk'} \delta n_{k'\alpha'} .
\label{eq:dressn0}
\end{equation}
Thus, the dressed current density operator takes the form
\begin{equation}
{\bf j} = \sum_{k\alpha} g_{k\alpha} {\bf v}_{k}\, \delta\tilde{n}_{k\alpha}
\label{eq:dresscurrent1}
\end{equation}
where
\begin{equation}
\delta\tilde{n}_{k\alpha} = \delta n_{k\alpha}
- \frac{\partial n_{0}}{\partial \epsilon_{k}} \sum_{k'\alpha'}
f^{\alpha\alpha'}_{kk'} \delta n_{k'\alpha'}
\label{eq:dresschange}
\end{equation}
is the difference between the true electron distribution and
the equilibrium distribution in the presence of Fermi
liquid interactions.

In a superconductor, there are two types of electrons: those that compose
the condensate of ground state pairs and those that form the Bogoliubov
quasiparticles.  Hence the equilibrium distribution function has both a
condensate term and a quasiparticle term:
\begin{equation}
n_{0} = n^{C}_{0} + n^{Q}_{0} .
\label{eq:twotypes}
\end{equation}
The equilibrium distribution of electrons in the condensate is just given
by the coherence factor $v_{k}^{2}$ so its derivative is
\begin{equation}
\frac{\partial n^{C}_{0}}{\partial \epsilon_{k}} = \frac{\partial}
{\partial \epsilon_{k}} \left[ \frac{1}{2} \left( 1 - \frac{\epsilon_{k}}{E_{k}}
\right) \right] = - \frac{\Delta_{k}^{2}}{2 E_{k}^{3}} .
\label{eq:conddist}
\end{equation}
In the presence of impurities,
the equilibrium quasiparticle distribution is given by the convolution of
the quasiparticle spectral function (a Lorentzian of width $\Gamma_{0}$ about
$E_{k}$) and the Fermi function $n_{F}$.  Multiplying this by the probability
the quasiparticle is an electron minus the probability it is a hole,
$(u_{k}^{2}-v_{k}^{2})$, we obtain the equilibrium distribution of electrons
that form the quasiparticles.  In the $T\ll\Gamma_{0}$
limit, its derivative is given by
\begin{eqnarray}
\frac{\partial n^{Q}_{0}}{\partial \epsilon_{k}}
&=& \frac{\partial}{\partial \epsilon_{k}} \left[ \frac{\epsilon_{k}}{E_{k}}
\int_{-\infty}^{\infty} \!\!\!\!\!\! d\omega
\frac{\Gamma_{0}/\pi}{(\omega - E_{k})^{2} + \Gamma_{0}^{2}} n_{F}(\omega)
\right] \nonumber \\
&=& - \frac{\epsilon_{k}^{2}}{E_{k}^{2}}
\frac{\Gamma_{0}/\pi}{E_{k}^{2} + \Gamma_{0}^{2}}
+ \frac{\Delta_{k}^{2}}{\pi E_{k}^{3}} \arctan \left( \frac{\Gamma_{0}}{E_{k}}
\right) .
\label{eq:quasidist}
\end{eqnarray}
Making use of these expressions and evaluating the dressed current
density operator,
\begin{equation}
{\bf j} = \sum_{k\alpha} g_{k\alpha} {\bf v}_{k} \left[
\delta n_{k\alpha}
- \left( \frac{\partial n^{C}_{0}}{\partial \epsilon_{k}}
+ \frac{\partial n^{Q}_{0}}{\partial \epsilon_{k}} \right)
\sum_{k'\alpha'} f^{\alpha\alpha'}_{kk'} \delta n_{k'\alpha'} \right]
\label{eq:dresscurrent}
\end{equation}
for the appropriate coupling parameters and velocities, the Fermi liquid
renormalizations of the electrical, thermal, and spin currents can
be computed.


\begin{references}
\bibitem{lee93} P. A. Lee, Phys. Rev. Lett. {\bf 71}, 1887 (1993)
\bibitem{hir93} P. J. Hirschfeld, W. O. Putikka, and D. J. Scalapino,
	Phys. Rev. Lett. {\bf 71}, 3705 (1993)
\bibitem{hir94} P. J. Hirschfeld, W. O. Putikka, and D. J. Scalapino,
	Phys. Rev. B {\bf 50}, 10250 (1994)
\bibitem{hir96} P. J. Hirschfeld and W. O. Putikka,
	Phys. Rev. Lett. {\bf 77}, 3909 (1996)
\bibitem{gra96} M. J. Graf, S-K. Yip, J. A. Sauls, and D. Rainer,
	Phys. Rev. B {\bf 53}, 15147 (1996)
\bibitem{sen98} T. Senthil, M. P. A. Fisher, L. Balents, and C. Nayak,
	cond-mat/9808001
\bibitem{bal94} A. V. Balatsky, A. Rosengren, and B. L. Altshuler,
	Phys. Rev. Lett. {\bf 73}, 720 (1994)
\bibitem{lee97} P. A. Lee and X. G. Wen, Phys. Rev. Lett. {\bf 78},
	4111 (1997)
\bibitem{mil98} A. J. Millis, S. M. Girvin, L. B. Ioffe, and A. I. Larkin,
	J. Phys. Chem. Solids {\bf 59}, 1742 (1998)
\bibitem{xu95} D. Xu, S. K. Yip, and J. A. Sauls, Phys. Rev. B {\bf 51},
	16233 (1995)
\bibitem{hos98} A. Hosseini, R. Harris, S. Kamal, P. Preston, R. Liang,
	W. N. Hardy, and D. A. Bonn, cond-mat/9811041
\bibitem{tai97} L. Taillefer, B. Lussier, R. Gagnon, K. Behnia, and
	H. Aubin, Phys. Rev. Lett. {\bf 79}, 483 (1997)
\bibitem{chi99} M. Chiao, R. W. Hill, C. Lupien, B Popic, R. Gagnon, and
	L. Taillefer, cond-mat/9810323
\bibitem{bon96} D. A. Bonn, S. Kamal, A. Bonakdarpour, R. Liang, W. N. Hardy,
	C. C. Holmes, D. N. Basov, and T. Timusk, Czech. J. Phys. {\bf 46},
	3195 (1996)
\bibitem{mes99} J. Mesot, M. R. Norman, H. Ding, M. Randeria, J. C. Campuzano,
	A. Paramekanti, H. M. Fretwell, A. Kaminski, T. Takeuchi, T. Yokoya,
	T. Sato, T. Takahashi, T. Mochiku, and K. Kadowaki, cond-mat/9812377
\bibitem{gor85} L. P. Gorkov and P. A. Kalugin, Pis'ma Zh. Eksp. Teor.
	Fiz. {\bf 41}, 208 (1985) [JETP Lett. {\bf 41}, 253 (1985)]
\bibitem{ner95} A. A. Nersesyan, A. M. Tsvelik, and F. Wenger, Nucl. Phys.
	{\bf B438}, 561 (1995)
\bibitem{sen98A} T. Senthil and M. P. A. Fisher, cond-mat/9810238
\bibitem{amb65} V. Ambegaokar and A. Griffin, Phys. Rev. {\bf 137},
	A1151 (1965)
\bibitem{amb64} V. Ambegaokar and L. Tewordt, Phys. Rev. {\bf 134},
	A805 (1964)
\bibitem{sch64} J. R. Schrieffer, {\it Theory of Superconductivity}
	(Addison-Wesley Publishing Company, Reading, MA, 1964)
\bibitem{mah90} G. D. Mahan, {\it Many-Particle Physics\/} (Plenum
	Press, New York, 1990)
\bibitem{mon87} H. Monien, K. Scharnberg, and D. Walker, Solid State Commun.
	{\bf 63}, 263 (1987)
\bibitem{hir88} P. J. Hirschfeld, P. W\"olfle, and D. Einzel, Phys. Rev. B
	{\bf 37}, 83 (1988)
\bibitem{mor96} J. Moreno and P. Coleman, cond-mat/9603079
\bibitem{chi99A} M. Chiao, P. Lambert, R. W. Hill, C. Lupien, R. Gagnon,
	L. Taillefer, and P. Fournier, cond-mat/9910367
\bibitem{wal96} S. F. Lee, D. C. Morgan, R. J. Ormeno, D. M. Broun,
	R. A. Doyle, and J. R. Waldram, Phys. Rev. Lett. {\bf 77}, 735 (1996)
\bibitem{leg65} A. J. Leggett, Phys. Rev. {\bf 140}, A1869 (1965)
\bibitem{lar63} A. I. Larkin and A. B. Migdal, Zh. Eksp. Teor. Fiz.
	{\bf 44}, 1703 (1963) [JETP {\bf 17}, 1146 (1963)]
\bibitem{lar64} A. I. Larkin, Zh. Eksp. Teor. Fiz. {\bf 46}, 2188 (1964)
	[JETP {\bf 19}, 1478 (1964)]
\bibitem{pin66} D. Pines and P. Nozieres, {\it The Theory of Quantum Liquids\/}
	(Addison-Wesley Publishing Company, Reading, MA, 1966)
\end{references}
\end{document}